\documentclass[10pt,journal,compsoc]{IEEEtran}
\ifCLASSOPTIONcompsoc
  \usepackage[nocompress]{cite}
\else
  \usepackage{cite}
\fi
\ifCLASSINFOpdf
\else
\fi
\usepackage[dvipsnames]{xcolor}
\usepackage[linesnumbered,lined,commentsnumbered]{algorithm2e}
\usepackage{mathabx}
\usepackage[colorlinks=true,linkcolor=black,anchorcolor=black,citecolor=black,filecolor=black,menucolor=black,runcolor=black,urlcolor=black]{hyperref}

\ifpdf%                                % if we use pdflatex
  \pdfoutput=1\relax                   % create PDFs from pdfLaTeX
  \usepackage{graphicx}                % allow us to embed graphics files
  \DeclareGraphicsExtensions{.pdf,.png,.jpg,.jpeg} % for pdflatex we expect .pdf, .png, or .jpg files
\else%                                 % else we use pure latex
  \usepackage{graphicx}                % allow us to embed graphics files
  \DeclareGraphicsExtensions{.eps}     % for pure latex we expect eps files
\fi%

\graphicspath{{figures/}{pictures/}{images/}{./}} % where to search for the images

\usepackage{microtype}                 % use micro-typography (slightly more compact, better to read)
\PassOptionsToPackage{warn}{textcomp}  % to address font issues with \textrightarrow
\usepackage{textcomp}                  % use better special symbols
\usepackage{mathptmx}                  % use matching math font
\usepackage{times}                     % we use Times as the main font
\usepackage{cite}                      % needed to automatically sort the references
\usepackage{tabu}                      % only used for the table example
\usepackage{booktabs}                  % only used for the table example
\usepackage[bottom]{footmisc}
\usepackage{stfloats}
\usepackage[normalem]{ulem}
\usepackage{flushend} % even column
\usepackage{wrapfig}
\usepackage{enumitem}
\setitemize{noitemsep,topsep=0pt,parsep=0pt,partopsep=0pt}

\usepackage{supertabular}

\usepackage{colortbl} % to color a row
\usepackage{multirow}
\definecolor{green}{RGB}{3, 128, 0}
\definecolor{yellow}{HTML}{FFF2CC}
\definecolor{lightGreen}{HTML}{E2F0D9}
\definecolor{visBlue}{HTML}{000000}
\definecolor{mlOrange}{HTML}{000000}
\newcommand*\rot{\rotatebox{90}}

\newcommand{\eg}{\emph{e.g.}}
\newcommand{\ie}{\emph{i.e.}}
\newcommand{\paperNum}{88}
\newcommand{\MLforVIS}{{\textcolor{mlOrange}{ML}4\textcolor{visBlue}{VIS}}}
\newcommand{\sixProc}{\textit{\revision{Data Processing4VIS}, \revision{Data-VIS Mapping}, Insight Communication, Style Imitation, VIS Interaction, \revision{VIS Reading}, and \revision{User Profiling}}}

\newcommand{\qianwen}[1]{\textcolor{black}{#1}}
\newcommand{\revision}[1]{\textcolor{black}{#1}}

\begin{document}
% \title{Applying \color{mlOrange} Machine Learning \color{black}Advances to \color{visBlue} Data Visualization\color{black}: A Survey on \color{mlOrange}ML\color{black}4\color{visBlue}VIS}
\title{A Survey on ML4VIS: Applying Machine Learning Advances to Data Visualization}

\author{Qianwen~Wang,
        Zhutian~Chen,
        Yong~Wang,
        and~Huamin~Qu% <-this % stops a space
\IEEEcompsocitemizethanks{
\IEEEcompsocthanksitem Qianwen Wang is with Harvard University.\\
E-mail: \url{qianwen_wang@hms.harvard.edu}
\IEEEcompsocthanksitem Zhutian Chen is with the University of California San Diego.\\
E-mail: \url{zhutian@ucsd.edu}
\IEEEcompsocthanksitem Yong Wang is with Singapore Management University.\\
E-mail: \url{yongwang@smu.edu.sg}
\IEEEcompsocthanksitem Huamin Qu is with Hong Kong University of Science and Technology.\\
E-mail: \url{huamin@cse.ust.hk}% <-this % stops an unwanted space
\IEEEcompsocthanksitem Part of this work was done when the first three coauthors were with Hong Kong University of Science and Technology.
}
\thanks{\textcopyright 2021 IEEE. Personal use of this material is permitted.
  Permission from IEEE must be obtained for all other uses, in any current or future 
  media, including reprinting/republishing this material for advertising or promotional 
  purposes, creating new collective works, for resale or redistribution to servers or 
  lists, or reuse of any copyrighted component of this work in other works.\\
  DOI: \href{<https://doi.ieeecomputersociety.org/10.1109/TVCG.2021.3106142>}{10.1109/TVCG.2021.3106142}}}


\markboth{A Survey on ML4VIS}%
{Shell \MakeLowercase{\textit{et al.}}: Bare Demo of IEEEtran.cls for Computer Society Journals}



\IEEEtitleabstractindextext{%
\begin{abstract}
Inspired by the great success of machine learning (ML), researchers have applied ML techniques to visualizations to achieve a better design, development, and evaluation of visualizations.
This branch of studies, known as ML4VIS, is gaining increasing research attention in recent years.
To successfully adapt ML techniques for visualizations, a structured understanding of the integration of ML4VIS is needed.
In this paper, we systematically survey \paperNum{} ML4VIS studies, aiming to answer two motivating questions: \textit{``what visualization processes can be assisted by ML?''} and \textit{``how ML techniques can be used to solve visualization problems?''}
This survey reveals seven main processes where the employment of ML techniques can benefit visualizations: \sixProc.
The seven processes are related to existing visualization theoretical models in an ML4VIS pipeline, aiming to illuminate the role of ML-assisted visualization in general visualizations.
Meanwhile, the seven processes are mapped into main learning tasks in ML to align the capabilities of ML with the needs in visualization. 
Current practices and future opportunities of ML4VIS are discussed in the context of the ML4VIS pipeline and the ML-VIS mapping.
While more studies are still needed in the area of ML4VIS, we hope this paper can provide a stepping-stone for future exploration.
% A web-based interactive browser of this survey is available at \texttt{\url{https://ml4vis.github.io}}.
\end{abstract}

}

\maketitle

\IEEEdisplaynontitleabstractindextext

% \fi
\IEEEpeerreviewmaketitle

\IEEEraisesectionheading{\section{Introduction}\label{sec:introduction}}
\nocite{sips2009selecting}
\nocite{gotz2009behavior}
\nocite{savva2011revision}
\nocite{VizDeck2012}
\nocite{steichen2013user}
\nocite{brown2014finding}
\nocite{Lalle2015curve}
\nocite{dereck2014}
\nocite{ddevaluation2015}
\nocite{vizrec2016}
\nocite{aupetit2016sepme}
\nocite{siegel2016figureseer}
\nocite{Kembhavi2016diagram}
\nocite{autosummarygeneration}
\nocite{pezzotti2016approximated}
\nocite{poco2017extracting}
\nocite{what2018kwon} %% 2019
\nocite{bylinskii2017learning}
\nocite{saha2017see}
\nocite{graphbytsne}
\nocite{poco2017reverse}
\nocite{jung2017chartsense}
\nocite{bylinskii2017understanding}
\nocite{AMLApproachforsss}
\nocite{siddiqui2018shapesearch}
\nocite{gramazio2017analysis}
\nocite{Draco2019} %% 2019
\nocite{generativemodel2015tvcg} %% 2019
\nocite{wang2017perception} %% 2017
\nocite{evaluatingGP} %% 2019
\nocite{luo2018deepeyekeyword} %% 29
\nocite{milo2018next}
\nocite{JUMP}
\nocite{kahou2017figureqa} %% 2018
\nocite{luo2018deepeye}
\nocite{FanH2018fast}
\nocite{chegini2018interactive}
\nocite{kafle2018dvqa}
\nocite{kim2018dynamic}
\nocite{battle2018beagle}
\nocite{data2vis2019CGA} % 2019
\nocite{haleem2019evaluating} % 2019
\nocite{madan2018synthetically}
\nocite{yu2019flowsense}
\nocite{he2019insitunet}
%%%%% For yong
\nocite{chen2019lassonet}
\nocite{han2019tsr}
\nocite{chen2019generativemap}
\nocite{kwon2019deep}
\nocite{wang2019deepdrawing}
\nocite{han2020flownet}
\nocite{wall2019markov}
\nocite{fujiwara2019incremental}
\nocite{fu2019vis_assessment}
\nocite{porter2019deep}
\nocite{jo2019disentangled}
\nocite{ma2018scatternet}
\nocite{wang2019datashot}
\nocite{cui2019text}
\nocite{chen2019towards}
\nocite{wang2019deeporgannet}
\nocite{colorCrafting2020TVCG}
\nocite{huang2019natural}
\nocite{hong2019dnn}
\nocite{fan2019KDE}
\nocite{ottley2019follow}
\nocite{abbas2019clustme}
\nocite{KasselR19Online}
\nocite{vizML2019CHI}
\nocite{fan2019personalized}
\nocite{kafle2020answering}
\nocite{mohammed2020continuous}
\nocite{zhang2020viscode}
\nocite{wu2020mobilevisfixer}
\nocite{tang2020plotthread}
\nocite{qian2020retrieve}
\nocite{wang2020vc}
\nocite{oppermann2020vizcommender}
\nocite{fosco2020predicting}
\nocite{giovannangeli2020toward}
\nocite{liu2020autocaption}
\nocite{luo2020interactive}
\nocite{lekschas2020peax}
\nocite{zhao2020iconate}
\nocite{lai2020automatic}
\nocite{kim2020answering}
\nocite{lu2020exploring}
\nocite{zhou2020table2charts}

\IEEEPARstart{D}{ata} visualization (\textbf{\textcolor{visBlue}{VIS}}), which maps data into visual presentations (\eg, position, color), is a powerful approach to uncover hidden insights and communicate compelling stories in data.
However, achieving an effective visualization is typically challenging, requiring a large amount of human effort and having a high reliance on expertise, such as graphic design, user experience design, and data analysis.
% including but not limited to graphic design, data analysis, and user experience.

% Recent years have witnessed remarkable progress in machine learning (ML) techniques.
Machine Learning (\textbf{\textcolor{mlOrange}{ML}}), on the other hand, provides a practical opportunity to relieve the reliance on experts in visualization.
By automatically learning knowledge from data, ML enables task completion without explicit instructions from humans.
\revision{The application of ML techniques can benefit a variety of visualization-related problems.
In this paper, we define visualization-related problems as problems that are related to the process of creating, interacting with, and evaluating visualizations.}
We refer to studies that \textit{apply ML techniques to solve visualization-related problems using the knowledge extracted from data}
as \textbf{\MLforVIS} in this paper.
ML4VIS studies can be traced back to 1986~\cite{mackinlay1986automating} and have been reignited by the recent advances in ML.
A series of ML4VIS studies are emerging,
% Ongoing advances in ML leads to the emergence of ML4VIS studies.
% Just in IEEE VIS 2019, more than a dozen of ML4VIS studies are presented, 
covering a wide range of visualization problems (\eg, graph layout~\cite{wang2019deepdrawing}, visualization assessment~\cite{fu2019vis_assessment}) and ML techniques (\eg, variational auto-encoders~\cite{kwon2019deep}, Mask-RCNN~\cite{chen2019towards}). 
% ML4VIS promise a more convenient creation, a more efficient interaction, and a more accurate analysis of visualization.
% These ML4VIS studies can effectively lower the barrier of visualization for non-experts. 
% These ML4VIS studies directly utilize the knowledge extracted from data and can effectively lower the barrier of visualization.
% effective improve the quality of visualizations and the user experience of interacting.
% For example, DeepEye~\cite{luo2018deepeye, luo2018deepeyekeyword}, a data visualization system, enables the automatic creation of meaningful visualizations for a given dataset without requiring any action from users.

% (why difficult & what is the gap， why a survey is needed)
While ML4VIS generates a stream of new opportunities, it also poses a series of challenges.
First, it is not always clear, among various problems related to visualization, which one can be significantly improved by existing ML techniques and which one still requires a high level of human intervention.
% Rough applying ML techniques may only impose the drawbacks of ML (\eg, uncertainty, inexplainability) on visualizations without any reward.
Roughly applying ML techniques to unsuitable visualization problems may only impose the drawbacks of ML (\eg, uncertainty, inexplainability) and threaten the validity of a visualization without bringing any benefit.
% -----------
% “the lack of interpretability and repeatability of ML-generated results, which may limit their application.”
% Meanwhile, there is a plethora of ML techniques, which are suitable for different problems.
Second, given a visualization-related task, selecting a proper ML technique and employing necessary adaptation are crucial yet challenging.
There are a plethora of ML techniques, most of which are exclusively suitable for a certain type of problems. 
For example, to automatically interpret infographics, 
% ML techniques developed for natural-image-understanding are inspiring but ML techniques developed for speech recognition are relatively irrelevant.
ML techniques developed for natural-image-understanding are more helpful and relevant than ML techniques developed for speech recognition.
% cannot be directly applied~\cite{Kembhavi2016diagram,chen2019towards}.
% A comprehensive understanding about the ML technique and the visualization problem is required to tailor these techniques for the visualization problem.
%=
% Third, the lack of interpretability in ML can limit its application in visualization.
% Users may lose trust on an ML4VIS work when the ML-generated result, such as a recommended visualization, is at odds with the users' expectation and cannot be explained.
% While uninterpretability is an inherent limitation of ML and may not be solved in the near future, 
% a better design, validation, and explanation of ML4VIS can help alleviate this issue.
% An comprehensive validation and explanation is required so that users can better understand and trust an ML4VIS work~\cite{}.
% that users can understand and trust.
% =
% Moreover, many ML techniques cannot be directly applied to visualization problems and require certain adaptation.
% For example, natural-image-understanding has been extensively studied in computer vision, but applying these techniques to charts and diagrams requires considerable adaptation~\cite{Kembhavi2016diagram,chen2019towards}.
% Therefore, a better understanding of the relationship between ML techniques and visualization problems can shed light on the area of ML4VIS and benefit the visualization community.
Therefore, the success of ML4VIS hinges on a better understanding of both the visualization and the ML, as well as the integration between the two.

\revision{
Researchers have contributed a series of frameworks and surveys about the integration between visualization and ML, but they mainly investigate how visualization can assist ML in data analysis~\cite{keim2015bridging, keim2008visual,Endert2017ML2VIS, yuan2020survey, chatzimparmpas2020state, chatzimparmpas2020survey} rather than how ML can be used to solve visualization-related problems.}
% For example, in the widely-accepted visual analytics (VA) pipeline proposed by Keim et al.~\cite{keim2008visual}, ML is only considered as a data processing component. 
For example, in a state-of-the-art report, Endert et al.~\cite{Endert2017ML2VIS} depicted a closer integration between visualization and ML techniques through interactive ML.
However, the goal of this integration is still to facilitate data analysis in different application domains (\eg, text analytics, multimedia analytics). 
Some recent surveys have reviewed studies on automatic visualization creation~\cite{zhu2020survey, vartak2017towards}, but these surveys consider both ML-based and non-ML-based methods.
More importantly, they only focus on one specific visualization-related problem, \ie, visualization creation. 
The opportunities for applying ML techniques to various visualization-related problems (\eg, visualization creation, visualization assessment) have not been fully discussed in previous studies.
To the best of our knowledge, this paper is the first published survey on ML4VIS.

% Advances in machine learning are enabling
% Insipred by the , researchers in the visualization community
% "This research combine" (define)
% "the ultimate goal is to "

% "Ongoing advances in AI technologies will continually generate 
% a stream of challenges and opportunities . 
% While such developments will require ongoing studies 
% and sufficient vigilance, we also see value in 
% developing reusable guidelines 
% that can be shared, refined, and debated"

% "Despite these efforts, more general solutions that blend machine learning and VA still do not exist. Yet, it is these more general tools that are needed to deal successfully with real-world challenges [21, 48]. Aiming at a more general understanding of how to integrate algorithmic and visual components, a wide variety of theoretical VA models and frameworks have been proposed [12, 20, 34, 47, 48]. These models, however, often focus on high-level, abstract views, and fail to successfully characterize how a strong interplay between algorithms and visualizations would be realized and exploited."

% (what did we do)
In this work, we systematically reviewed the literature
in related fields including data visualization, human-computer interaction (HCI), 
ML, and data mining to investigate how ML techniques are employed and adapted for visualization. 
Build upon existing visualization theoretical models, we proposed in an ML4VIS pipeline, which reveals seven main processes that can benefit from employing ML techniques. 
The seven visualization processes (\ie, \sixProc) are mapped into the main learning tasks in ML to bridge the needs in visualization with the capabilities of ML.
% we related this survey to existing theoretical visualization models and incorporate the six process into an ML4VIS pipeline. 
We discuss the current practices and future opportunities of ML4VIS in the context of the ML4VIS pipeline and the ML-VIS mapping.
% We illustrated how previous ML4VIS studies can fit in this ML4VIS pipeline and discussed research opportunities derived from our analysis. 
Specifically, our contributions are:
\begin{itemize}[leftmargin=9pt]
    \item A codification of \paperNum{} ML4VIS studies into \textbf{seven processes} in \textcolor{visBlue}{visualization} that can benefit from the employment of \textcolor{mlOrange}{ML} techniques (\autoref{sec:six_process});
    \item An \textbf{\MLforVIS{} pipeline} that integrates the seven visualization processes with existing visualization models (\autoref{sec:ml4vis_pipeline}) and an \textbf{\textcolor{mlOrange}{ML}-\textcolor{visBlue}{VIS} mapping} between the seven visualization processes and an ML taxonomy (\autoref{sec:map_ml}).
    \item A set of \textbf{research challenges and opportunities} derived from our analysis (\autoref{sec:opportunities}).
\end{itemize}
% - A codifcation of over 150 AI-related design recommendations collected from academic and industry sources into a set of 18 generally applicable design guidelines for human- AI interaction (see Table 1).
% • A systematic validation of the 18 guidelines through multiple rounds of iteration and testing.
We hope this paper can provide a stepping stone for further exploration in the area of \MLforVIS.
An interactive browser of this survey is available at 
\textbf{\texttt{\url{https://ml4vis.github.io}}}.
% \qianwen{
% The remainder of the paper is organized as below.
% We first review related surveys and visualization models (\autoref{sec:related_work}) and then introduces the survey methodology used in this paper (\authoref{sec:survey_method}).
% \authoref{sec:six_process} summarizes six visualization processes that benefit from ML techniques in our survey. The six visualization processes are connected using a ML4VIS pipeline that are proposed in \authoref{sec:ml4vis_pipeline}.
% }

% "We investigate specific implementations of visual analysis systems integrating DR, and analyze ways that other machine learning methods have been combined with DR. 
% Summarizing the results in a “human in the loop” process model provides a general lens for the evaluation of visual interactive DR systems. We apply the proposed model to study and classify several systems previously described in the literature, and to derive future research opportunities."

\section{Related work}
\label{sec:related_work}
This paper mainly relates to two streams of literature: surveys that aim to guide ML-assisted design and theoretical models that summarize the process in visualization.

% \subsection{Guiding the Designs of ML4HCI}
\subsection{Guiding ML-assisted Design}

As various ML capabilities (\eg, recommendation, interaction prediction)
are integrated into user interfaces,
many surveys and reviews have been conducted to guide the application of ML techniques in the design of user interfaces~\cite{yang2018mapping, Amershi2019guidelines,Horvitz1999principles}. 
% Visualizations, which can effectively improve user perception and promote user engagement, play an important role in the filed of HCI.

These surveys on general ML-assisted designs have contributed various guidelines and taxonomies that also inform the ML4VIS studies.
Efforts to review these guidelines can be traced back to 1999, when Horvitz~\cite{Horvitz1999principles} reviewed and outlined 12 guidelines for coupling automated services with direct user manipulation. 
These guidelines were later extended by Amershi et al.~\cite{Amershi2019guidelines},
who distilled 18 design guidelines for human-AI interaction from over 150 AI-related design recommendations.
The majority of these guidelines, such as \textit{``learn from user behaviors''}, \textit{``convey the consequence of user interactions''},
can also be applied to the design of ML-assisted interactive visualizations.
Apart from design guidelines, 
% taxonomies are also contributed to
taxonomies can also provide a better understanding of ML techniques through a design perspective.
Yang et al.~\cite{yang2018mapping} analyzed more than 2000 HCI publications using topic modeling and summarized four channels through which ML advances can provide value to users: self-understanding, contextual awareness, optimization, and utility-capability. The four channels can also help people envision new ways of ML techniques to improve visualization.
While these surveys shed some light on the study of ML4VIS,
visualizations have certain unique characteristics that require specified investigations.
For example, a precise understanding of and an efficient interaction with data are required in visualization, but they are rarely discussed in the general design of user interfaces.

To guide ML-assisted visualization designs, Saket et al.~\cite{saket2018beyond} compared and categorized existing approaches for learning visualization design principles based on the learning method and the input features.
The authors then described a research agenda for deriving visualization design principles using ML techniques.
Even though this study provides a helpful and insightful agenda, it focuses on the creation of visualization and makes little investigation about user interaction and perception, which are important in interactive visualizations.
In this study, we present a survey on ML4VIS that covers various visualization processes.

\subsection{Theoretical Models in Visualization}
Theoretical models (\eg, workflows, pipelines) 
have become increasingly important in designing, developing, and evaluating visualizations.
A variety of theoretical models have been proposed to depict different processes in visualization~\cite{munzner2009nested,keim2008visual,van2005value}.
For example, Card~\cite{card1999readings} presented a visualization workflow to describe the process of creating and interacting with visualizations.
Munzner~\cite{munzner2009nested} proposed a nested model for the design and validation of visualization. \looseness=-1

\revision{
With the growing usage of ML in visualization,
more and more studies have included ML techniques as a component in the theoretical models of visualization.
Keim et al.~\cite{keim2008visual} presented the first effort to form a general VA pipeline, where ML is included as a data processing module.
Follow-up studies proposed theoretical models tailored for different analysis scenarios and ML techniques~\cite{sacha2016visual,sacha2014knowledge,el2018visual}.
For example, El-Assady et al.~\cite{el2018visual} extended the general VA pipeline to encompass a new visual analytics paradigm, \ie, speculative execution.
Sacha et al.~\cite{sacha2016visual} introduced a process model to describe the visual interaction with dimension reduction, a widely-used ML technique.
% In these theoretical models, ML and visualization work together to assist users in data analysis.
Endert et al.~\cite{Endert2017ML2VIS} reviewed theoretical models that embed ML techniques into visual analytics and listed opportunities for future research.
In these studies, ML is integrated with visualization techniques to facilitate the analysis of complex data.
The capabilities of ML in enabling more effective visualizations still remain unclear.
}

Recently, there is a growing research trend on modeling the role of visualization in the design, development, and evaluation of ML~\cite{yuan2020survey, chatzimparmpas2020survey, chatzimparmpas2020state, keim2015bridging, Sacha2019vis4ml}.
For example, a Dagstuhl Seminar~\cite{keim2015bridging} identified a series of research opportunities in combining ML with visualization.
Sacha et al.~\cite{Sacha2019vis4ml} reinterpreted the traditional VA pipeline and proposed a VIS4ML ontology to encompass ML development workflows into VA workflows.
Spinner et al.~\cite{spinner2019explainer} presented a VA framework for interactive and explainable ML.
A research gap exists in summarizing how ML can contribute to visualization. 

Contrary to those prior theoretical models, we investigated how ML techniques can be used to benefit the creation, evaluation, and interaction of visualizations.
We present an ML4VIS pipeline based on our survey of \paperNum{} papers.
While more studies are still needed in the field of ML4VIS, 
we hope this pipeline can help map out the current landscape and
inform future opportunities. 
% \section{Landscape of ML4VIS}
\section{Survey Methodology}
\label{sec:survey_method}
% In this study, we aimed to map out the landscape of ML4VIS studies, identify opportunities for future research.
To provide an understanding of the current ML4VIS studies, we conducted an analysis of \paperNum{} related papers in the field of VIS, HCI, ML, and data mining \& management.
Each paper is coded based on the \textbf{\textcolor{visBlue}{VIS process}} and the \textbf{\textcolor{mlOrange}{ML task}}.
Even though automatic analysis methods (\eg, topic modeling) are popular among recent survey papers~\cite{yang2018mapping,sacha2016visual,LiuWCDOEJK2019bridging}, automatic methods usually only provide a high-level understanding of a large number of papers (\eg, more than 2000 \cite{yang2018mapping}).
\revision{Manual analysis is still needed to extract details for a deep analysis~\cite{sacha2016visual, LiuWCDOEJK2019bridging}.
% ML4VIS, as a new emerging topic, has a relatively small body of literature that can be manually analyzed.
Therefore, to ensure a thorough analysis, we conducted a manual analysis for the collected papers.}

% We analyzed these papers through a keyword-based selection, a manual validation, and a manual encoding.

% A complete list of the surveyed papers and their codes are available in the supplementary material.

% Even though automatic analysis methods (\eg, topic modeling) are popular among recent suvey papers~\cite{yang2018mapping,sacha2016visual,LiuWCDOEJK2019bridging}, automatic methods only provide a high-level understanding of a large number of papers (\eg, more than 2000 \cite{yang2018mapping}).
% Manual analysis is still needed to extract details for a deep analysis~\cite{sacha2016visual, LiuWCDOEJK2019bridging}.
% ML4VIS, as a new emerging topic, has a relatively small body of literature that can be manually analyzed.
% Therefore, to ensure a thorough analysis, we conducted a manual analysis for the collected papers.

\renewcommand{\arraystretch}{1.5} %increase the vertical padding for this table
\begin{table}[]
    \centering
    \caption{Relevant venues}
    \begin{tabular}{m{3cm}|m{4cm}}
    \toprule
          VIS \newline (visualization) & IEEE VIS (InfoVis, VAST, SciVis), EuroVis, PacificVis, TVCG\\
          \hline
          HCI  (human-computer  \newline interaction) & CHI, UIST, IUI \\
          \hline
          DMM  (data mining and management) &  KDD, SIGMOD, ICDE, VLDB\\
          \hline
          ML  (machine learning) &  CVPR, ECCV, ICML, NeurIPS, ICCV, AAAI, IJCAI, ICLR \\
         \bottomrule
    \end{tabular}
    \vspace{0.5em}
    
    \label{tab:conferences}
\end{table}
\renewcommand{\arraystretch}{1} %set the vertical padding back to 1.1

\begin{figure}
    \centering
    \includegraphics[width=\linewidth]{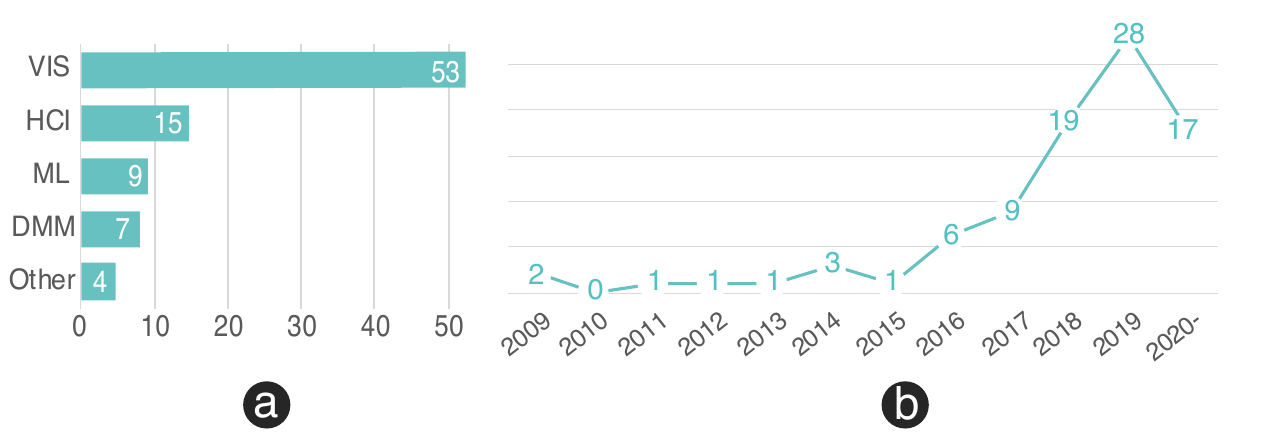}
    \caption{(a) The number of ML4VIS studies published in different fields: VIS, HCI, DMM (data mining and management), ML;(b) The number of ML4VIS studies over time.}
    \label{fig:survey_statistics}
\end{figure}

% \subsection{Paper Selection}
\subsection{Survey Scope}
We began our analysis by assembling a corpus of ML4VIS papers.
We collected all research articles published between 2016 Jan to 2020 Oct in main journals and conferences in the field of VIS, HCI, ML, and data mining \& management \qianwen{by directly accessing the venues}, as shown in \autoref{tab:conferences}. 
\qianwen{
We selected impactful journals and conferences in corresponding fields mainly according to Google Scholar Top Publications~\cite{GoogleScholar}. 
}
% This time range is based on two main considerations: 
% 1) an increasing number of research papers on applying ML techniques to visualization appear in the latest five years, 
% 2) given that ML techniques are progressing rapidly, the latest research is more important for informing readers of the latest trend in ML4VIS. 
The time range was chosen based on two main considerations: 
a) this time range has covered the majority of the state-of-art ML4VIS studies
% known to the authors 
and is manageable to conduct a detailed manual analysis;
b) given that ML techniques are progressing rapidly, the latest research can provide better guidance for follow-up studies. 
Moreover, papers published before this time range will be included later by going through the representative references of the collected papers.
Following the practice in~\cite{yuan2020survey}, we mainly checked the title and abstract of each paper to strike a balance between efficiency and accuracy.
We went through the full manuscript only when the title and abstract cannot provide clear information.
During this process, we paid special attention to a set of ML-related keywords 
(\eg, \textit{``learning''}, \textit{``machine''}, \textit{``training''}, \textit{``AI''}, \textit{``automation''}, \textit{``CNN''}, \textit{``LSTM''})
and visualization-related keywords 
(\eg, \textit{``visualization''}, \textit{``infographic''}, \textit{``diagrams''}, \textit{``charts''}).
After such a round of paper selection and filtering, we obtained 259 papers.

\begin{table}[]
\caption{All surveyed papers and their codes.}
\vspace{-4mm}
    \centering
     \includegraphics[width=0.9\linewidth]{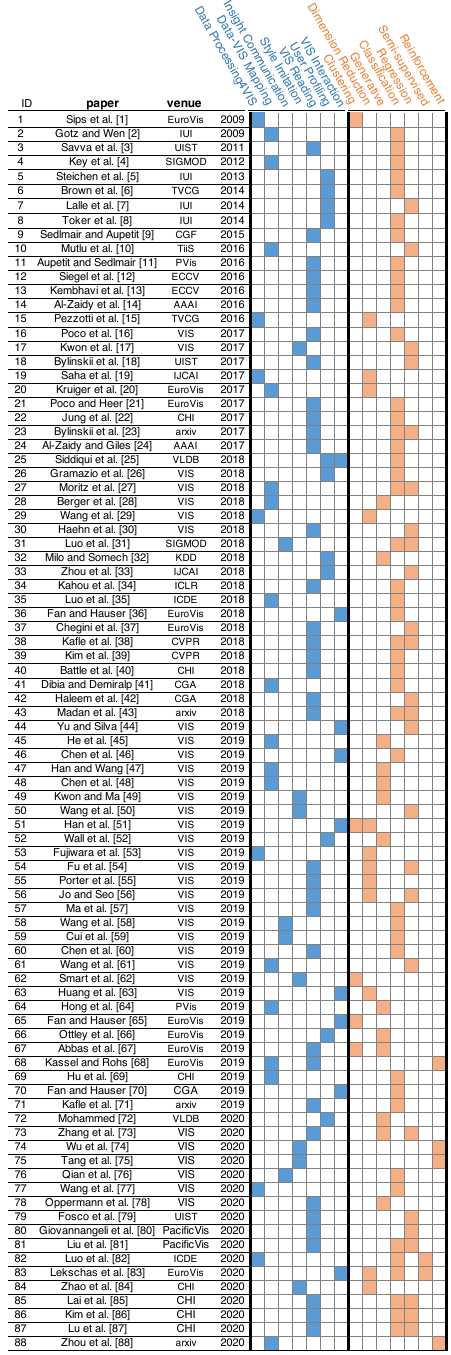}

    \label{tab:all_papers}
\end{table}

\revision{
We further carefully checked the 259 papers using the following criteria.
First, since this survey aims to understand the current practices of employing ML techniques for visualizations, we focused on papers that contribute novel techniques and applications.  
Theory, evaluation, and dataset papers were excluded.
Then, we validated the three terms, \textcolor{mlOrange}{\textbf{ML}}, \textcolor{visBlue}{\textbf{VIS}}, and \textbf{4 (for)}, respectively.
% 
% validate ml
To validate the term \textcolor{mlOrange}{\textbf{ML}}, a paper should employ ML techniques to learn from training data rather than totally depend on human defined rules.
This criterion requires that a training or optimization process is conducted on a collection of sample data.
For example, Voyager~\cite{Wongsuphasawat2017voyager} was excluded from the corpus because it recommends charts using a list of permitted channels and marks defined by human experts.
% 
% validate vis
To validate the term \textcolor{visBlue}{\textbf{VIS}}, ML should be used for visualization-related problems, including the creation, interaction, and evaluation of visualizations.
In other words, we consider information visualization, scientific visualization, and visual analytics except those studies that purely use ML techniques for data processing.
This criterion distinguishes ML4VIS studies from studies that purely use ML to analyze data.
For example, NNVA~\cite{hazarika2019nnva} employs ML techniques to analyze the parameter sensitivity of yeast cell polarization simulations. In this study, visualizations are merely used to represent the sensitivity analysis that are generated by ML. Therefore, NNVA is not an ML4VIS study and was eliminated from the corpus.
% 
% validate for
To validate the term \textbf{4 (for)}, a paper should contribute a novel application or an improvement of existing ML techniques.
This criterion helps focus on insightful applications and better scope the range of this survey.
For example, a large number of papers that directly use the existing dimension reduction methods to visualize high-dimensional data were not included since they fail to improve these techniques or apply these techniques to a new problem.
But the paper~\cite{graphbytsne} that employs t-SNE to layout graph and the paper~\cite{wang2017perception} that improves LDA to maximize the perceived separation of humans were included.
Following the practice in other surveys~\cite{guo2020survey, yuan2020survey}, we \qianwen{conducted one-round reference-based search and} further went through the related work of these papers.
To focus on the latest development in ML4VIS and provide insights for follow-up studies, we did not \revision{include} papers that were published before 2000.
We collected 12 referenced papers that satisfy the aforementioned criteria but appeared earlier than 2016 or on other venues.
Our final corpus included \paperNum{} relevant papers.
A summary of the survey statistics is shown in \autoref{fig:survey_statistics}.
}

\subsection{Paper Coding}
The paper analysis \revision{consisted} of three phases.
% Inspired by \cite{amini2015understanding}, we conducted both close coding and open coding.
In the first phase, we extracted a brief description for each paper, including the targeted visualization problems, the employed ML models, and the collected training data. 
In the second phase, we coded the collected papers from a visualization perspective and analyzed what visualization problems are solved by ML techniques.
We categorized these visualization problems based on the visualization processes that these problems are related to.
We aimed to answer the question \textbf{``what \textcolor{visBlue}{visualization processes} can be automated by ML?''} from this encoding.
We referred to the visualization processes in existing visualization pipelines \cite{van2005value, card1999readings} and modified them accordingly to better fit the context of ML4VIS.
Each of the three authors independently coded the content of 60 papers to ensure that each paper was coded by at least two authors.
\qianwen{
Note that, at this stage, we did not have unified names for all the visualization processes.
The three authors then discussed their own codes about visualization processes with all authors.
% In the first phase, three co-author independently used an open-coding approach to characterize the content of 20 (x\%) collected papers. 
During weekly discussion, we iteratively adapted, split, and refined the codes multiple times until there was no further disagreement and finally arrived at seven main visualization processes.
}
More details of the seven visualization processes are explained in \autoref{sec:six_process}.
Meanwhile, the seven processes form an ML4VIS pipeline, whose differences from existing visualization pipelines are discussed in \autoref{sec:ml4vis_pipeline}.
In the third phase, we sought to characterize \textbf{``how \textcolor{mlOrange}{ML} can be used to solve visualization problems?''} 
% We followed a widely-used ML taxonomy~\cite{nicolas2015scala} and analyzed the ML4VIS studies around the type of learning tasks. Each of the three co-authors coded around 1/3 of the collected papers. 
As new ML models constantly emerge and one ML model is usually applied to solve various problems, we found that it is not practical to code ML4VIS studies based on the used ML models.
Therefore, we answered this question by coding the types of learning tasks.
% Therefore, we answered this question by coding the formats of training data and the types of learning tasks, which are the two key factors in choosing a suitable ML model.
% We referred to \cite{nicolas2015scala, murphy2012machine} for the coding of learning tasks and employed a bottom-up approach for the coding of data formats.
As with phase two,
each of the three authors coded 60 papers to ensure each paper was at least coded by two authors. 
We employed a bottom-up approach and referred to the learning tasks in \cite{nicolas2015scala, murphy2012machine}.
The learning tasks are summarized and discussed in \autoref{sec:map_ml}.
% Since an ML4VIS paper sometimes involves more than one learning task, we only coded the most important learning task.
The codes of all collected papers are summarized in \autoref{tab:all_papers}.
\revision{For those papers that appeared both at IEEE VIS conference and IEEE TVCG, we listed their venue as VIS.}
% The VIS codes are explained in Table~\ref{table:six_process} and introduced in details in \autoref{sec:six_process}.
% The ML codes are explained and discussed in \autoref{sec:map_ml}. 
Most of the ML4VIS papers used one ML technique to solve one major visualization problem. But it is also possible that an ML4VIS paper involves more than one learning task and one visualization process, so a few papers are labeled with more than one ML or VIS code.

\begin{table*}
  \centering
  \caption{Seven main visualization processes that benefit from employing ML techniques.}

  % \begin{tabular}{m{0.3cm}|m{2cm}|m{3cm}|m{5.5cm}|m{5.5cm}} 
    \begin{tabular}{@{\hspace{-2pt}}m{0.3cm}|m{2cm}|>{\raggedright}m{3cm}| @{\hspace{0.1\tabcolsep}} m{11cm}} 

    \cmidrule[1.5pt]{1-4}
  & Process 
  & Description 
  & \begin{tabular}{>{\raggedright}m{4cm}|m{6cm}} 
    % Main Purposes of Employing ML 
    \revision{Problems Solved by ML}
   & Representative Examples   \\
\end{tabular}
  % & Main Goals of Existing ML4VIS studies 
  % & Representative Examples  
  \\  
      
  \cmidrule[1pt]{1-4}

  \rot{DATA} & \revision{Data \newline Processing4VIS} 
   & raw data is transformed into a format that better suits the following visualization processes
   & \begin{tabular}{>{\raggedright}m{4cm}|m{6cm}} 
        improve the efficiency of data processing
        & ML cleans data for a specified visualization by asking questions to users \cite{luo2020interactive}
         \\ \hline 
        enhance perception of generated visualizations 
        & ML transforms high dimension data to 2D and maximizes the human-perceived separation among classes\cite{wang2017perception}
        \\  
    \end{tabular}
   \\ 
   \cmidrule[1pt]{1-4}

   & \revision{Data-VIS \newline Mapping}
  %  & the already processed data is mapped into visual representations.
%   & visualizations are created to present the already processed data
& \revision{data fields are mapped into visual channels}
   & 
   \begin{tabular}{>{\raggedright}m{4cm}|m{6cm}} 
        generate suitable visualizations 
        & ML generates Vega-Lite visualization specifications for a given dataset \cite{data2vis2019CGA} 
        \\ \hline 
        improve the efficiency of visualization creation 
        &
        % ML enables flexible insuite visualization of ensemble simulations\cite{he2019insitunet} 
        ML synthesizes intermediate density map images between given density maps without storing and visualizing data of all time steps \cite{chen2019generativemap}
        \\  
    \end{tabular}
  \\ 
  % \toprule
  \cmidrule[0.5pt]{2-4}

  \multirow{3}{*}{\rot{VIS}} & Insight \newline Communication 
    &  \revision{insights are embedded in visualizations to be effectively communicated}
%   &  visualizations are created to communicate the given insights
   & \begin{tabular}{>{\raggedright}m{4cm}|m{6cm}} 
        interpret insights 
        & ML recognizes entities from the user-provided text insights \cite{cui2019text}
        \\ \hline 
        generate suitable visualizations
        % map insights to visualizations 
        & ML predicts suitable visualization specifications for communicating given insights \cite{luo2018deepeyekeyword}
        \\  
    \end{tabular} 
  %   & \begin{tabular}{m{5.5cm}} 
  %     interpret insights \\ \hline 
  %     mapping insights to visual representations \\  
  % \end{tabular}  
   \\ 
   \cmidrule[0.5pt]{2-4}

  & Style Imitation 
  %  & create visualizations that share similar visual properties with given visual designs
%   & visualizations are created with similar visual properties of the given examples
    & \revision{styles are extracted from the given examples and applied to the created visualization}
   & \begin{tabular}{>{\raggedright}m{4cm}|m{6cm}} 
       imitate the color selection
       & ML generates color ramps by imitating designers' practices in choosing colors \cite{colorCrafting2020TVCG} 
        \\ \hline
       imitate the layouts 
       & ML generates graph drawings with similar layout styles of the given examples
       \cite{wang2019deepdrawing} 
       \\ 
    \end{tabular} 
  %   & \begin{tabular}{m{5.5cm}} 
  %     \cite{colorCrafting2020TVCG} \\ \hline
  %     \cite{wang2019deepdrawing} \\ 
  %  \end{tabular}  
  \\ 
   \cmidrule[1pt]{1-4}
   
   \multirow{2}{*}{\rot{USER~~~~~~~~~~~~~~~~~~~~~~~~}} 
   
   \color{red}
   & VIS Interaction 
   & users interact with a visualization and transformed it into a new stage
   through user actions
   & \begin{tabular}{>{\raggedright}m{4cm}|m{6cm}} 
          refine the result of the current interaction
          % & ML refines the points that are selected by drawing a lasso in 3d point cloud \cite{chen2019lassonet}
          & ML refines the selected points of a lasso selection in 3D point clouds \cite{chen2019lassonet}
          \\ \hline
        %   predict next-step interactions
        %   & ML predicts a user's next click in a scatterplot~\cite{ottley2019follow}
        %   \\ \hline
          \qianwen{understand natural interaction}
          & ML interprets the given text descriptions and extracts the corresponding visual patterns~\cite{siddiqui2018shapesearch}
          \\  
          
        %   learn about users
        %   & ML predicts a user's task performance and personality based on mouse interaction~\cite{brown2014finding}
        %   \\ 
          
      \end{tabular}
      \\ 
  \cmidrule[0.5pt]{2-4}
  
  & \revision{ User \newline Profiling }
   & \revision{ user actions with visualizations are logged and analyzed to better understand users }
   & \begin{tabular}{>{\raggedright}m{4cm}|m{6cm}} 
        %   refine the result of the current interaction
          % & ML refines the points that are selected by drawing a lasso in 3d point cloud \cite{chen2019lassonet}
        %   & ML refines the selected points of a lasso selection in 3D point clouds \cite{chen2019lassonet}
        %   \\ \hline
        
        %   predict next-step actions
            \revision{predict user behavior}
          & ML predicts a user's next click in a scatter plot~\cite{ottley2019follow}
          \\ \hline
        %   understand natural interactions 
        %   & ML interprets the given text descriptions and extracts the corresponding visual patterns~\cite{siddiqui2018shapesearch}
        %   \\ \hline 
          
          \revision{predict user characteristics}
          & ML predicts a user's task performance and personality based on mouse interaction~\cite{brown2014finding}
          \\ 
          
      \end{tabular}
      \\ 
  \cmidrule[0.5pt]{2-4}
  
  \color{black}

   & \revision{VIS \newline Reading} 
  %  & users observe the appearance of a visualization, read the encoded data, and understand the underlying insights. 
  & users read visualizations and obtain useful information 
  % \qianwen{not sure}
    &\begin{tabular}{>{\raggedright}m{4cm}|m{6cm}} 
      extract content 
      & ML extracts an extensible visualization template from a timeline infographics~\cite{chen2019towards}
      \\ \hline 
      interpret content 
      & ML answers questions about a given bar chart visualization~\cite{kafle2018dvqa}
    %   ML analyzes a large amount of real-world online visualizations\cite{battle2018beagle}
      \\ \hline
      \revision{estimate human perception}
      % & ML predicts a user's learning curve with a specific visualization \cite{Lalle2015curve}
      & ML predicts the distribution of users' attention on an infographic~\cite{bylinskii2017learning}
      \\
    \end{tabular}
    % &\begin{tabular}{m{5.5cm}} 
    %   content extraction \\ \hline 
    %   content understanding \\ \hline
    %   user modeling \\
    % \end{tabular}
     \\ 

    \cmidrule[1.5pt]{1-4}
  \end{tabular}
  \vspace{0.5em}
  
  \label{table:six_process}
\end{table*}

\section{Benefit from ML: Seven VIS Processes}
\label{sec:six_process}
% We define ML4VIS as \textit{``an automaic process, use ML techniques, 
% either the input or the output contains visualizations (a vega configuration or an image).
% ''},

In this section, we introduce seven visualization processes emerged from our literature review that are benefiting from ML techniques, as shown in \autoref{table:six_process}.
% Each process represents a transition from certain inputs to outputs.
We first explain the transition from inputs to outputs at each visualization process,
and then discuss the \revision{problems solved by} ML at each process with representative examples.
% The validation of ML techniques in the seven visualization processes is also summarized.
% and then present an ML4VIS pipeline. 
% Each process indicates a transition from certain inputs to outputs that can be facilitated by ML techniques. 
% We explain how ML techniques can assist these processes with representative examples from the survey.
% Note that these seven visualization processes are not necessarily mutually exclusive and are often incorporated with each other in ML4VIS studies. \looseness=-1
% The seven processes fall into three dimensions: \colorbox{yellow}{raw materials}, \colorbox{lightBlue}{visualizations}, \colorbox{lightGreen}{users} 

\subsection{Data Processing4VIS}

\begin{figure}[]
  \centering
  \includegraphics[width=\linewidth]{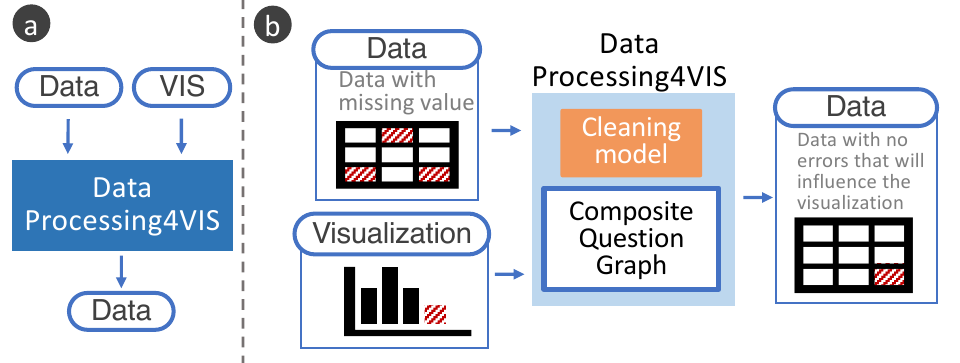}
  \caption{(a) An illustration of the \emph{\revision{Data Processing4VIS}}; (b) An example of employing ML in \emph{\revision{Data Processing4VIS}}: Luo et al.~\cite{luo2020interactive}. An ML model (orange box) is trained to detect data errors and propose cleaning options.}
  \label{fig:data_clean}
\end{figure}

\noindent
\textbf{Process Description:} 
In \emph{\revision{Data Processing4VIS}} (\autoref{fig:data_clean}(a)), raw data is transformed into a format that better suits the following visualization processes.
% To scope this survey and highlight the focus on ML4VIS, 
% To highlight the focus on ML4VIS, we distinguish \emph{\revision{Data Processing4VIS}} from the general data processing used in previous VA studies by emphasizing the property of \emph{VIS-driven}.
% Specifically, 
Contrary to the general-purpose data processing, \emph{\revision{Data Processing4VIS}} refers to a process that is tightly related to and exclusively designed for the context of visualization, such as enabling efficient visualization creation and enhancing human perception.

\vspace{0.7em}
\noindent
\textbf{\revision{Problems Solved by ML:}}  
% \emph{\revision{Data Processing4VIS}} transforms raw data for the specific visualization-related purposes.
% The reviewed studies commonly conduct three data processing tasks: transformation\cite{wang2017perception}, sampling, cleaning\cite{luo2020interactive}.
% For data transformed, wang et al.~\cite{wang2017perception}
In \emph{\revision{Data Processing4VIS}}, ML techniques demonstrate the ability to make raw data better satisfy certain visualization-related purposes, 
including enabling efficient visualization creation 
and enhancing human perception.

 \noindent $\blacktriangleright$ \emph{Enabling efficient visualization creation.}
  By narrowing down the purpose of data processing to the creation of specific visualizations, 
  the employment of ML techniques can improve the efficiency of data processing.
  For example, data cleaning, an inevitable step in data processing, is often treated as an independent step and completed before the visualization creation.
%   Instead,
%   Luo et al.~\cite{luo2020interactive} 
%   related the data cleaning processes with visualization creations
%   and used ML to enable a VIS-driven data processing
%   that allows the user to interactively clean the data during the creation of a visualization.
 Instead, Luo et al.~\cite{luo2020interactive} 
  related the data cleaning processes with visualization creations
  and used ML techniques to help users interactively clean the data after the visualization creation.
  % The progressive data cleaning in \cite{luo2020interactive} 
  % is VIS-driven since that the error detections and cleaning options are specified for the visualization the user choose.
  % \ie, it requires user feedback and only works in interactive visualizations.
  \autoref{fig:data_clean}(b) illustrates an overview of \cite{luo2020interactive}.
  Users first specify a visualization and create it using the uncleaned data.
  A trained model then detects errors and generates cleaning options for the data underlying the visualization. 
  These cleaning options are provided to users in the form of yes/no questions.
  % Visualizations are first created using uncleaned data. 
  % Questions about how to clean data for the created visualization are then generated.
  Based on users' answers, the underlying data is cleaned and the visualization is updated.
  This progressive data cleaning enables a more flexible and efficient creation of visualization, 
  since data does not need to be totally cleaned before creating visualizations.

\noindent $\blacktriangleright$ \emph{Enhancing human perception.} 
  Meanwhile, ML techniques also demonstrate the ability to enhance user perception of the created visualizations in \emph{\revision{Data Processing4VIS}}.
  % Data transformation is another important step in data processing.
  % For example, dimension reduction is an widely used ML technique in general data analysis as well as a common data transformation for 2D visualization. 
  % Since it enables a interpretable visualization of high dimensional data, it is also widely used in transforming the data for visualization. 
  For example, dimension reduction (DR) is a widely used data processing method for visualizing high dimensional data.
  Previous studies often directly employ the general DR methods proposed in ML field without considering the needs in visualizations.
  \revision{To better present data in scatter plots, Wang et al.~\cite{wang2017perception} proposed a supervised DR method that takes the perceptual capabilities of humans into account.
  The modified DR method mimics human perception in cluster separation and successfully maximizes the visually perceived cluster separation in 2D projections.} \looseness=-1

% According to our survey, we identify two visualization related purposes: enabling 

% \noindent
% \textbf{Validation}
% Since most tasks in data processing have ground truth or widely-accepted metrics, they can be evalauted through quantitative evaluation~\cite{luo2020interactive}.
% However, when the results involve subjective factors such as user perception, user studies are still required for validating the methods.

\subsection{\revision{Data-VIS Mapping}}

\begin{figure}[]
  \centering
  \includegraphics[width=\linewidth]{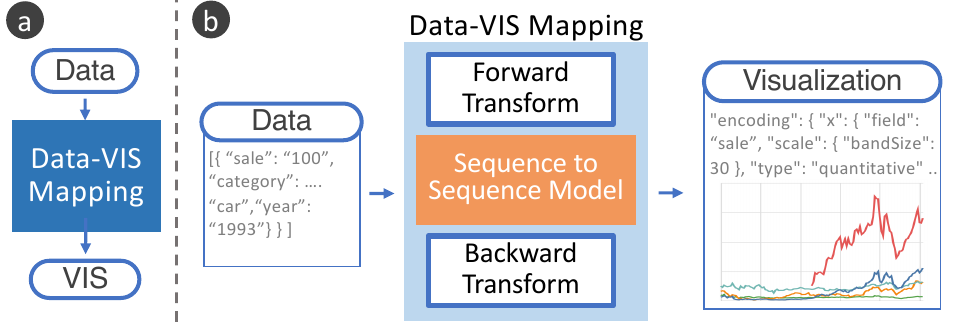}
  \caption{(a) An illustration of the \emph{\revision{Data-VIS Mapping}} process. (b) A representative example of employing ML in \emph{\revision{Data-VIS Mapping}}: Data2Vis~\cite{data2vis2019CGA}.
  % The pipeline of Data2Vis demonstrates an example of ML4VIS in \emph{\revision{Data-VIS Mapping}}. 
  An ML model (orange box) is trained to map JSON-encoded datasets into Vega-lite visualization specifications.}
  \label{fig:data2vis}
\end{figure}

\noindent
\textbf{Process Description:}  
In \emph{\revision{Data-VIS Mapping}} (\autoref{fig:data2vis}(a)), 
% the processed data is transformed into appropriate visual representations in order to help people better understand and analyze the data.
% This process is achieved through a systematic mapping from data values to graphic marks.
\revision{the values of data fields are mapped into the visual channels of graphic marks.
Appropriate visual mapping is needed in this process to help people better understand and analyze the visualized data.}
% This transition can be completed either by directly generating bitmap images or by creating the visualization specifications
%  which can be either bitmap images or text configurations like Vega.  
% ML techniques can facilitate this process by generating appropriate visualizations in a more efficient way.
Such a mapping is usually manually specified using code or authoring tools, which results in steep learning curves and makes data visualization inaccessible to general users. 

% The creation of data visualizations often requires manual specification through code or interactions, which results in steep learning curves and makes data visualization inaccessible to general users. 

\vspace{0.7em}
\noindent
\textbf{\revision{Problems Solved by ML:}}   
In the collected ML4VIS studies,
% literature review,
ML techniques mainly facilitate \emph{\revision{Data-VIS Mapping}} by recommending suitable visual representations and by improving the efficiency of visualization creation. \looseness=-1

 \noindent $\blacktriangleright$ \emph{Recommending visualizations.}
  A set of recent research has shown that ML can be used to automatically recommend suitable visual representations.
  %  facilitate the automated recommendation of visualizations.
  For example, DeepEye~\cite{luo2018deepeye} combines supervised ML techniques with expert rules to automatically recommend good visualizations for users.
  Given a dataset, all the possible visualizations are enumerated and classified as ``good'' or ``bad'' by a binary decision tree classifier.
  All the ``good'' visualizations are then ranked and provided to users.
  % DeepEye~\cite{luo2018deepeye} combines supervised ML techniques with expert rules to classify visualizations as ``good'' or ``bad'' and further rank multiple visualization choices to provide users with top-k good visualizations for a given dataset. 
  % Also, deep neural network models have been popularly used to facilitate the process of \emph{\revision{Data-VIS Mapping}}.
  Meanwhile, the great success of deep learning in various domains promotes its application in \emph{\revision{Data-VIS Mapping}}.
%   By collecting a large visualization dataset with data features and design choices,
  VizML~\cite{vizML2019CHI} collects a large visualization dataset and trains a fully connected neural network to predict the top five design choices when creating visualizations for a specific dataset.
  Data2Vis~\cite{data2vis2019CGA} formalizes the process of \emph{\revision{Data-VIS Mapping}} as a sequence to sequence translation from the original data to the visualization specifications. 
  As illustrated in \autoref{fig:data2vis}(b),
  a sequence-to-sequence neural network learns the mapping between JSON-encoded datasets (source sequence) and Vega-lite visualization specifications (target sequence).
  % Such an end-to-end approach can effectively reduce the manual efforts in selecting data attributes and visual encodings, and specifying the mapping between data attributes and visual encoding variables.
  In these studies, ML automatically learns design choices from the training dataset and directly translates data to suitable visualizations. 
  As a result, the employment of ML can effectively reduce the manual efforts in selecting data attributes, 
  designing visual representations, 
  and specifying the mapping from data values to graphic marks.
  
 \noindent $\blacktriangleright$ \emph{Improving the efficiency of visualization creation.}
  % Another line of research focues on generating appropriate visualizations in an efficient way. 
  Another line of research focuses on improving the efficiency of creating visualizations.
   One direction is to improve the efficiency of creating dynamic visualizations.
  For instance, creating dynamic density maps usually leads to high computational cost and memory demand since a large amount of data needs to be recorded and then visualized at each time step. 
  To address this issue, 
  GenerativeMap~\cite{chen2019generativemap} proposes a generative model to synthesizes a series of density map images and show the dynamic evolution
  between two given density maps, thus relieving the burden of storing intermediate results.
  % Given two density maps,
  % the generative model synthesizes  between the two density maps.
  % It is even worse, if the dynamic evolution using density maps.
  % At each time step, a large amount of data needs to be recorded and then visualized through a pixel-based calculation.
  % To address this issue, GenerativeMap~\cite{chen2019generativemap} proposes a generative model to visualize the dynamic evolutions between density maps.
  % without the need to record data and render visualizations of all intermediate steps.
  Similarly, to visualize the dynamics in spatial-temporal data, TSR-TVD~\cite{han2019tsr} applies a recurrent generative network to learn the pattern in dynamic evolution and generate temporal super-resolution visualizations.
  In these studies, ML learns the patterns in dynamic evolution from training data and directly synthesize intermediate visualizations to reduce the needed storage and computation cost.
%   In addition to synthesizing intermediate steps between two visualizations,
%   ML can also be used to synthesize visualizations
Another interesting direction is to facilitate parameter exploration in visualization creation.
%   some researchers use ML to directly synthesize visualizations based on various input parameters (\emph{e.g.}, view points, transfer functions). 
%   without actually executing the expensive creating processes.
%   replace the rendering process by directly 
%   to enable people to explore synthesized visualizations under various input parameters 
%   directly synthesize visualizations based on parameters.
  For example, 
  Berger et al.~\cite{generativemodel2015tvcg} trained a Generative Adversarial Network (GAN) to synthesize visualizations 
  and guide users in transfer function editing by quantifying expected changes in the visualization.
  Similarly, InsituNet~\cite{he2019insitunet} introduces a deep learning based surrogate model to create visualizations of simulation data by learning the mapping from simulation parameters to visualization images. 
  These methods enable people to explore synthesized visualizations under various input parameters without actually executing the expensive creation processes.

  % For example, the visualization of volume data is a typical computationally intensive process and requires iterative modification of input parameters such as viewpoints and transfer functions.

  % To facilitate the exploration of transfer functions, 
  % The GAN connects the changes of transfer functions to the changes of output visualizations, thus guiding people in the creation of visualization. 
  % Another direction is to facilitate the parameter space exploration when creating visualizations~\cite{generativemodel2015tvcg, he2019insitunet, hong2019dnn}.

  % Another example is the visualization of volume data, which is a typical computationally intensive process.

  % In a traditional workflow, the visualization will be recalculated after the parameter modification, which further increases the computation complexity.

\subsection{Insight Communication}

\begin{figure}[]
  \centering
  \includegraphics[width=\linewidth]{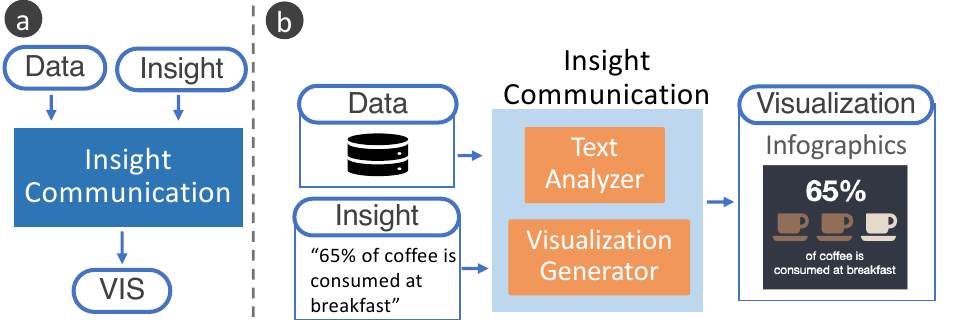}
  \caption{(a) An illustration of the process of \emph{Insight Communication}. (b) One representative example of ML4VIS in \emph{Insight Communication}: Text-to-Viz\cite{cui2019text}. A text insight is first analyzed by a text analyzer and then be transformed into expressive infographics by a visualization generator.}
  \label{fig:text2vis}
\end{figure}

\noindent
\textbf{Process Description:} 
In \emph{Insight Communication} (\autoref{fig:text2vis}(a)), insights are transformed into visualizations that can effectively communicate them.
As with many studies in the visualization community~\cite{north2006toward, chang2009defining}, we refer to the term \emph{Insight} as the knowledge about the data that is communicated through visualizations.
The main difference between \emph{Insight Communication} and \emph{\revision{Data-VIS Mapping}} is the availability of insights.
In \emph{\revision{Data-VIS Mapping}}, insights are hidden in the underlying data and need to be discovered by people through visual exploration.
In \emph{Insight Communication}, on the contrary, insights are already available to designers and need to be highlighted in the created visualizations. 
% Defferent from \emph{\revision{Data-VIS Mapping}}, where the insights are hidden in the underlying data and wait to be discovered by the analysts through visual exploration, \emph{Insight Communication} describes the visualization creation process where insights are already available to viusalization designers.
This difference is also denoted as exploratory versus explanatory visualization in prior studies~\cite{kim2019exploration}.
To better communicate insights, visualizations in \emph{Insight Communication} are usually more visually pleasing to improve user engagement and memorability.
% - input: data, insight
% - output: visualization (a configure like vega, or a bitmap image)
% - Usually design for explanatory data visualization.
% - contrary to present data, where patterns in the data is unknown, in this process, ML is used to assist in the communication of an insight, a story that the creators already found

% The key problem in \emph{Insight Communication} is to choose the approriate visualization designs, including data mapping, visual encoding, and view composition, that can effective and accurately convey specific insights.

\vspace{0.7em}
\noindent
\textbf{\revision{Problems Solved by ML:}}  
In \emph{Insight Communication}, existing studies have successfully applied ML techniques to interpret insights and to build a mapping from insights to suitable visualizations.

\noindent $\blacktriangleright$ \emph{Interpreting insights.}
  % The insights can be either provided by the designer through natural language or automaicaly generated using data mining techniques.
  The insights can be either provided by the designers or automatically generated from the data.
  For example, Text-to-Viz~\cite{cui2019text} allows users to provide insights through natural language statements such as \textit{``40\% of USA freshwater is for agriculture''} (\autoref{fig:text2vis}(b)).
  A text analyzer (a supervised CNN+CRF model) is trained to extract entities from the sentence and understand the provided insights.
%   Denoting insights as strong manifestations of statistical properties (\eg, correlation, skewness),
DataShot~\cite{wang2019datashot} automatically identifies insights from tabular data and presents these identified insights to the users through infographics.
% Foresight~\cite{demiralp2017foresight} automatically identifies insights from large datasets and presents top-k insights to the users using visualizations.\looseness=-1

\noindent $\blacktriangleright$ \emph{Mapping insights into visualizations.}
ML techniques also demonstrate the capability to learn the mapping from insights to suitable visual representations.
  To facilitate this mapping, ML4VIS studies usually describe visual representations using specifications in a predefined design space to ensure a well-defined output space.
  % A predefined design space ensures a finite and well-defined output space, which facilitates the employment of ML techniques.
  Using the information from insights, ML can identify candidates for each dimension in the design space 
  and produce the final visualization from these valid combinations~\cite{wang2019datashot, cui2019text,luo2018deepeyekeyword}.
  For example, DeepEye~\cite{luo2018deepeyekeyword} employs declarative visualization languages similar to Vega-Lite to enable the creation of common visualizations (\eg, bar, pie charts).
  A decision tree model learns the mapping from insights and data characteristics to visualization specifications.
  Meanwhile, to generate visualizations with higher aesthetic values in \emph{Insight Communication}, researchers also contribute more advanced design spaces.
  For example, Text-to-Viz~\cite{cui2019text} summarizes four design-space dimensions of infographics--layout, description, graphic, and color--from discussion with design experts.
  A visualization generator synthesizes infographics by generating a tuple of values on these four dimensions.
  % Using the infomation extracted from insights, ML models identify candidates from each design dimension and produce the final visalization though ranking all valid combinations.
  % Wite a predefined design space, ML models can 

% \noindent
% \textbf{Validatian}
% Since the assessment of visualization is highly subjective, ML4VIS studies in \emph{Insight Communication} are usually evaluted through conducting in-lab user studies or providing a gallery of examples~\cite{wang2019datashot, cui2019text}.
% Using labeled preference data as ground truth, quantitative evaluation is also possible but is relatively more costly~\cite{luo2018deepeye}.

\begin{figure}[]
  \centering
  \includegraphics[width=\linewidth]{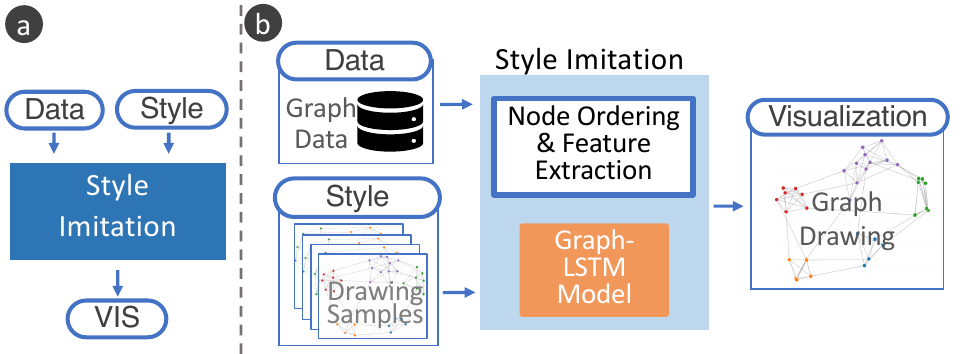}
  \caption{(a) An illustration of the process of \emph{Style Imitation}. (b) A representative example of ML4VIS in \emph{Style Imitation}: DeepDrawing~\cite{wang2019deepdrawing}. A Graph-LSTM model (orange box) learns the drawing styles from a set of samples and generates drawings of a similar style for the input data.}
  \label{fig:DeepDrawing}
%   \vspace{-1em}
\end{figure}

\subsection{Style Imitation}

\noindent
\textbf{Process Description:} 
In \emph{Style Imitation} (\autoref{fig:DeepDrawing}(a)), the styles of given visualizations examples are applied to create new visualizations. 
% We use the term \emph{style} to denote a set of variables that affect the aesthetic appearance of visualizations without influencing how data are encoded.
We use the term \emph{style} to denote a set of variables that affect the appearance of visualizations but are not directly mapped to data values.
Typical visualization styles include the color palette, the chart decoration, and the view layout and aspect ratio.
The main difference between \emph{Data2VIS Mapping} and \emph{Style Imitation} is whether the visual properties are mapped to data values.
For example, \emph{Data-VIS Mapping} decides whether data should be visualized as a scatter plot or a bar chart. In contrast, \emph{Style Imitation} extracts color palettes from given examples for creating a scatter plot.\looseness=-1

\vspace{0.7em}
\noindent
\textbf{\revision{Problems Solved by ML:}}   
% According to our survey, 
Even though the visualization style involves a large number of variables,
existing ML4VIS studies on \emph{Style Imitation} mainly apply ML techniques on two aspects, the imitation of color selection and the imitation of layout styles.

\noindent $\blacktriangleright$ \emph{{Imitating color selection.}}
  Color selection in visualizations is not only an aesthetic choice but also influences the effectiveness of information communication.
  Even though some helpful guidelines about colors are available, it is still challenging for novice visualization developers to use these loosely defined qualitative guidelines to generate high-quality color ramps.
  % For the imitation of color styles, a representative example is \emph{Color Crafting}~\cite{colorCrafting2020TVCG}, which assists users with quick creatation of effective color ramps. 
  % %that high-quality color ramp relies on designers' multiple years experiences and expertise and
  To address this issue,
  Color Crafter~\cite{colorCrafting2020TVCG} mimics the practices of professional designers and automatically generates high-quality color ramps. 
  The color style imitation is achieved by modeling the paths that expert-designed color ramps traverse through color space.
  The authors trained ML models in a corpus of 222 expert-designed color ramps that are summarized from popular visualizations, and used these models to generate effective color ramps for developers from one single seed color.
  % The design practices first are documented by assembling a corpus of 222 expert-designed color ramps from popular visualizations.
  % Clustering algorithms are then used to capture design practices by modeling the paths that these color ramps traverse through color space.
  % Color Crafter enables developers to create effective color ramps from a single seed color.
  % A collection of designer-crafted color ramps of popular visualizations. 
  % Color Crafter allows novice visualization developers to craft high-quality ramps.
  
\noindent $\blacktriangleright$ \emph{{Imitating layouts.}}
  % The other line of existing 
  A parallel line of research on \emph{Style Imitation} focuses on generating visualizations of similar layouts.
  
  ML techniques have been extensively employed to generate graph drawings that mimic the layout styles of given graph drawing examples.
  Kwon et al.~\cite{what2018kwon} presented an ML approach to facilitate large graph visualization by learning the topological similarity between large graphs.
  %%% V1 -- detailed version
%   Specifically, their approach will first visualize a set of large graphs and store their drawings. Given a new input large graph, their approach will assess the topology structure similarity between the input graph and the graphs with rendered drawings, and further directly show the drawing of a similar graph as a visualization preview for the input graph.
  %%% V2 --- short version
  Specifically, the method provides a quick overview of visualizations for an input large graph data by learning from the drawings with similar topology structures.
%
%
%   used machine learning models such as Support Vector Machine (SVM) to provide a quick overview of large graphs by using the rendered drawings of graphs with similar topology structures.
This approach is further extended in a recent study~\cite{kwon2019deep}, which learns the layout from examples using a deep generative model. The model is then used to provide users with an intuitive way to explore the layout design space of the input data.
Meanwhile, DeepDrawing~\cite{wang2019deepdrawing} trains a graph-LSTM to learn one specific layout style from graph drawing examples.
Instead of drawing a single graph in diverse layouts, the trained ML model in DeepDrawing
directly maps new input data into graph drawings that share similar layout styles with the training examples, as shown in \autoref{fig:DeepDrawing}(b).
  %  Apart from graph drawings, machine learning techniques have also been applied to style imitation of other visualizations, such as timeline inforgraphics~\cite{chen2019towards}.
  
  Apart from graph drawing, a few studies have explored imitating the layouts of other types of visualizations, including storyline visualization~\cite{tang2020plotthread} and mobile visualization~\cite{wu2020mobilevisfixer}. 
  For example,
  PlotThread~\cite{tang2020plotthread} trains a reinforcement learning agent to learn the layout of human-designed storyline visualizations.
%   Optimization-based methods create storyline layouts by optimizing a set of narrative constraints and aesthetic metrics, but they fail to consider the whole design space of storylines and   
%   While automatically-generated storylines can optimize, this AI agent learns to conduct a series of layout modifications to mimic the human designs from a trial-and-error processes.
  By imitating the layout from human-drawn examples, PlotThread is able to include more diverse narrative elements and create more expressive storylines compared with optimization-based methods.

% \noindent
% \textbf{Validation} Similar to \emph{\revision{Data-VIS Mapping}}, the assessment of using machine learning techniques to facilitate \emph{Visualization Style Imitation} also often involves both qualitative evaluations, such as use study~\cite{colorCrafting2020TVCG}, use cases~\cite{colorCrafting2020TVCG,generativemodel2019kwon}, example galleries~\cite{colorCrafting2020TVCG,wang2019deepdrawing,what2018kwon,generativemodel2019kwon,chen2019towards}, and quantitative evaluations, such as similarity metrics~\cite{what2018kwon,wang2019deepdrawing,generativemodel2019kwon}, time cost~\cite{what2018kwon,wang2019deepdrawing,generativemodel2019kwon} and accuracy~\cite{chen2019towards}.

%
%\noindent
%\textbf{Validation}

\begin{figure}[]
  \centering
  \includegraphics[width=\linewidth]{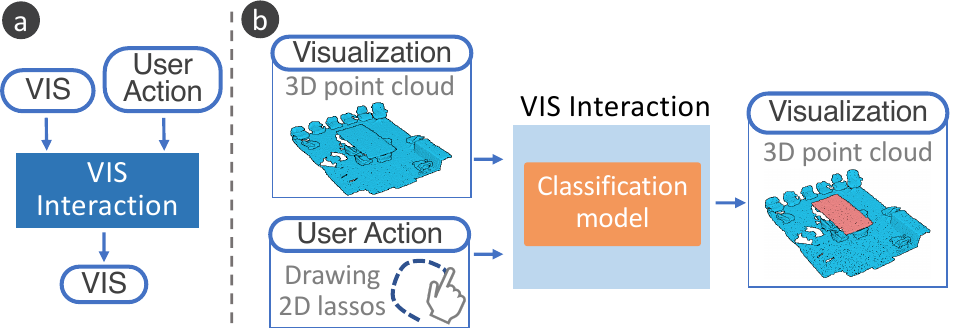}
  \caption{(a) An illustration of the \emph{VIS Interaction} Process. (b) A representative example of employing ML in \emph{VIS Interaction}: LassoNet~\cite{chen2019lassonet}. An ML model (orange box) predicts the user intended node selection in a 3D point cloud from a lasso drawn on 2D space.}
  \label{fig:lassoNet}
\end{figure}

\subsection{VIS Interaction}

\noindent
\textbf{Process Description:} 
In \emph{VIS Interaction} (\autoref{fig:lassoNet}(a)),
the appearance of a visualization is modified based on the user actions (\eg, zoom, filter).
% a visualization is transformed from to a new visualization through user interactions with the visualization (\eg, zoom, filter)
\emph{VIS Interaction} usually reflects user intentions in visual exploration and modifies the visualization by manipulating the view configurations, 
the visual mappings, or the underlying data~\cite{Dimara2020What}.

\vspace{0.7em}
\noindent
\textbf{\revision{Problems Solved by ML:}}  
In \emph{VIS Interaction}, previous studies have demonstrated the capabilities of
ML techniques to infer user intentions
and thus assist users in interactions.
According to our survey, the visualization problems solve by ML techniques in \emph{VIS Interaction} can be classified into two main groups,
refining interaction results and
understanding natural interactions.

\noindent $\blacktriangleright$ \emph{Refining interaction results.}
  ML techniques can refine the results of user interactions
  to achieve more accurate and efficient interactions.
  When interacting with a large number of visual elements, 
  an accurate and efficient selection is crucial but usually hard to achieve.
  Take the brushing in 2D space as an example.
  Traditional interactions often need to make a trade-off between interaction efficiency and accuracy, \eg, rectangular brushing is fast but inaccurate while the logical combination is accurate but slow.
%   are either fast but inaccurate (\eg, rectangular brushing) or accurate but slow (\eg, logical combinations of simple brushes).
  Fan and Hauser~\cite{FanH2018fast} exploited a CNN model to achieve both fast and accurate node selection in 2D scatter plots.
  The CNN estimates the intended node selection based on a simple line-brushing and data distribution in the visualization.
  Compared with selections in 2D space,
  node selection in 3D space is even more challenging: visualizing 3D points in 2D space easily causes occlusion; input devices such as touchscreens only operate in 2D space.
  To facilitate the node selection in the large-scale 3D point cloud, LassoNet~\cite{chen2019lassonet} uses a deep learning model to predict the node selection based on a lasso drawn on a 2D surface, the user's viewpoint, and characteristics of the point cloud.
  A pipeline of LassoNet is illustrated in \autoref{fig:lassoNet}(b).
  % These methods usually can help users interact with large-scale elements more easily.
  % For example, Chen et al.~\cite{chen2019lassonet} used a deep learning model to predict the selected points from a 3D point cloud based on a lasso drawn on a 2D surface.
  % Similarly, Fan and Hauser~\cite{FanH2018fast}
  % introduce a deep learning-based method to
  % facilitate brush selections on 2D scatter plots.
  % \qianwen{I have to completely rewrite the whole paragraph.}

 \noindent $\blacktriangleright$ \emph{Understanding natural interactions.}
  Natural interactions are interactions that humans naturally communicate through, such as gestures, natural language, and sketches~\cite{valli2008design}.
  ML techniques have demonstrated the capabilities to understand nature interactions and allow a more intuitive and convenient way to interact with visualization.
  % Compared with traditional visualization systems with WIMP (Windows, Icons, Menus, Pointer) user interfaces, systems with natural interfaces require less prior knowledge of the system functionality and thus usually have a more gentle learning curve.
  Given that the input natural interaction (\eg, sentence, sketch)
  can be fuzzy and uncertain,
  ML techniques help interpret the user's input
  and trigger dedicated functions.
  % Researchers have been exploring the ways to allow users to interact with visualizations naturally,
  % such as using natural language or sketch.
  For example, ShapeSearch~\cite{siddiqui2018shapesearch} allows users to search visual patterns from trend line visualizations using sketches and text descriptions.
  The authors trained a CRF model to recognize entities and understand natural interactions.
  FlowSense~\cite{yu2019flowsense} supports users in visual data exploration via natural language command, \eg, \emph{``highlight the selected cars in a parallel coordinates plot''}.
  While the language understanding is mainly accomplished by a non-ML method (semantic parsing) in FlowSense,
  a supervised learning model is trained to effectively resolve syntactic ambiguity in the language commands. \looseness=-1

% \color{green}
\subsection{User Profiling}

\begin{figure}[]
  \centering
  \includegraphics[width=\linewidth]{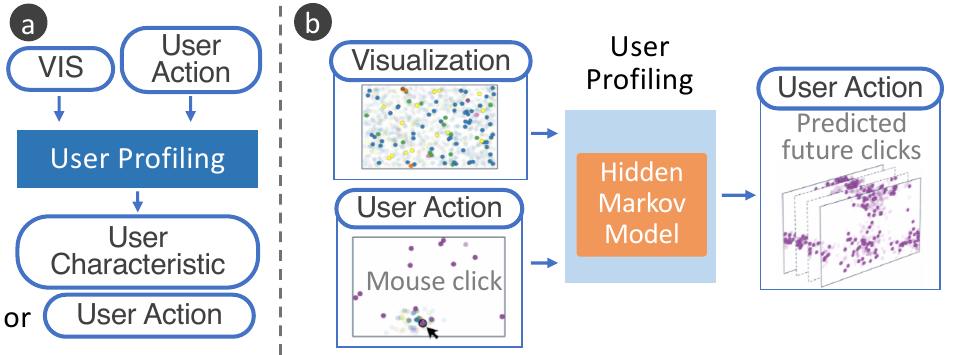}
  \caption{\revision{(a) An illustration of the \emph{User Profiling} process. (b) A representative example of employing ML in \emph{\revision{User Profiling}}: Ottley et al.\cite{ottley2019follow} employed a hidden Markov model (orange box) to predict user future actions based on their past actions and the visualization.} }
  \label{fig:user_understanding}
\end{figure}

\noindent
\textbf{Process Description:} 
In \emph{User Profiling} (\autoref{fig:user_understanding}(a)),
user actions with visualizations are logged and then analyzed in order to better understand users. 
% the appearance of a visualization is modified based on the user actions (\eg, zoom, filter).
% % a visualization is transformed from to a new visualization through user interactions with the visualization (\eg, zoom, filter)
% \emph{VIS Interaction} usually reflects user intentions in visual exploration and modifies the visualization by manipulating the view configurations, 
% the visual mappings, or the underlying data~\cite{Dimara2020What}. 

\vspace{0.7em}
\noindent
\textbf{\revision{Problems Solved by ML:}}  
In \emph{User Profiling}, previous studies have demonstrated the capabilities of ML models in modeling user behaviors and characteristics.

\noindent $\blacktriangleright$ \emph{Predicting user behaviors.}
  ML techniques can predict next-step interactions and facilitate complex visual explorations that consist of a series of interactions.
  Therefore, the employment of ML can ease the learning curve for users or 
  reduce the interaction latency in complex visual explorations.
%   reducing the computational cost for the computer.
  % which usually has a steep learning curve for users and a high computational cost for the computer.
  For example,
  to ease the learning curve, REACT~\cite{milo2018next} models the analysis context using a generic tree-based model, where the edges represent the user's actions and the nodes represent the system states. 
%   Given an analysis context, a context-similarity metric is utilized to fetch relevant contexts, which are then used to generate next-action suggestions to the user.
Given an analysis context, relevant contexts are first fetched using a context-similarity metric and then used to generate next-action suggestions for the user.
%   and generates next-action suggestions to users.
  % To assist users in tasks that require a series of interactions,
  % ML has been used to predict and recommend
  % the next move based on the user's current interaction.
  Another example is the study by Ottley et al.~\cite{ottley2019follow}.
  To reduce the interaction latency, Ottley et al. utilized a Markov model to predict the user's
  next mouse clicks in large scatter plots and fetch related data in advance. 
  % so that the system can make better preparation, \emph{e.g.}, prefetching related data.

\noindent $\blacktriangleright$ \emph{Predicting user characteristics.}
  Interactions with the visualization can reflect users' reasoning processes in visual exploration and even their own personal characteristics. ML can effectively analyze these interactions and help designers learn about the users, including their learning abilities~\cite{Lalle2015curve}, their cognitive abilities~\cite{steichen2013user}, their analysis goals~\cite{gramazio2017analysis,gotz2009behavior}, and even their personalities~\cite{brown2014finding}.
  For example, Gramaz et al. \cite{gramazio2017analysis} demonstrated that 
  the classification of mouse interactions can effectively infer the visual analysis tasks of cancer genomics experts. The ML classification can even expand current knowledge about the visualization tasks in cancer genomics.
  Brown et al.~\cite{brown2014finding} applied ML techniques to analyze users' mouse interactions in the visual exploration of finding Waldo. They found that ML can accurately predict the user task performance (\ie, the time used to find Waldo) and uncover user personalities (\eg, locus of control, extraversion, and neuroticism).
  Apart from mouse interactions, eye gaze data has also been treated as a type of interaction data and can help learn about users \cite{cooke2006mouse}.
  Lall{\'{e}} et al~\cite{Lalle2015curve},
  used ML techniques to predict a user's learning curve of a visualization based on the eye gaze data.
  By learning about users, visualization developers are able to go beyond the one-size-fits-all design and provide adaptive visualizations for different task needs, user abilities, and analysis stages.

\color{black}

\subsection{\revision{VIS Reading}}

\begin{figure}[]
  \centering
  \includegraphics[width=\linewidth]{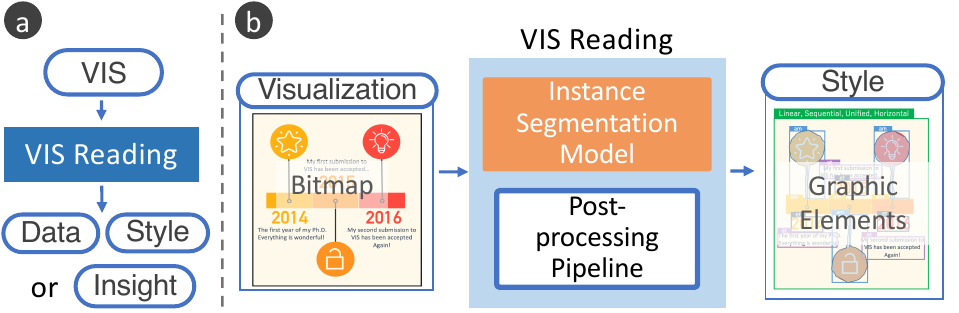}
  \caption{(a) An illustration of the \emph{\revision{VIS Reading}} process. (b) A representative example of employing ML in \emph{\revision{VIS Reading}}. Chen et al.\cite{chen2019towards} employed a deep learning model (orange box) to extract the graphical elements from a timeline infographics. }
  \label{fig:timeline}
\end{figure}

\noindent
\textbf{Process Description:} 
In \emph{\revision{VIS Reading}} (\autoref{fig:timeline}(a)), users observe the appearance of a visualization, read the encoded data, and understand the underlying information. 
ML techniques have been used in \emph{\revision{VIS Reading}} process to automatically \emph{``read''} the visualizations like humans, thus helping designers better understand users and enabling the analysis of a large corpus of visualizations.
% \qianwen{[what is the benefits of using ML]}
% Given a visualization as an input, an ML model can automatically extracts data, visual encoding, styles, and insights from the visualization.
% The input of this process is visualizations,
% usually in the form of bitmap image,
% and the output can be data, visual encodings, styles, and even insights.

% - use ML to interpret visualization in a manner that is similar to the way humans read and understand visualizations

\vspace{0.7em}
\noindent
\textbf{\revision{Problems Solved by ML:}}  
% To enable machines to read visualizations,
% ML techniques, especially the techniques from the field of \emph{Computer Vision},
% have been employed in the \emph{\revision{VIS Reading}} process.
In \emph{\revision{VIS Reading}}, the capabilities of ML techniques can be divided into three main groups:
extracting content,
understanding content,
and modeling users.
% extracting information from visualizations,
% perceiving visualizations in a manner similar to human visual system,
% and gaining high-level understanding from visualizations.

% simulating humans to perceive visualizations.
% Computer vision is an interdisciplinary scientific field that deals with how computers can gain high-level understanding from digital images or videos. From the perspective of engineering, it seeks to understand and automate tasks that the human visual system can do
% The information extract from
% a visualization can be used
% for the fofollowing tasks
% or modeling human's 

\noindent $\blacktriangleright$ \emph{Extracting content.}
    % ML can perform reverse engineering on visualizations
    % to extract data, visual encodings, graphical elements that 
    % can be used for applications such as indexing, restyling or retargeting, and summarizations for large-scale visualizations.
    ML techniques can directly extract content from visualizations in the form of bitmap images, such as 
    graphical elements~\cite{chen2019towards}, color palette~\cite{poco2017extracting}, and visual encodings~\cite{poco2017reverse}.
    The extracted content can be used for indexing, restyling, and reusing visualizations.
    % A typical example is ReVision~\cite{},
    % a system that extract the underlying data of charts.
    % In this system, a SVM model is trained to identify the chart types that is further used to infer the data.
    A typical example is \cite{poco2017reverse}, which contributes an end-to-end pipeline to extract Vega visualization specifications from standard visualizations such as line charts.
    A set of classification models are trained to identify chart types, localize text, and recognize graphic marks separately. 
    Compared with standard visualizations, infographics have more diverse appearances and are more difficult to interpret.
    Chen et al.~\cite{chen2019towards} utilized a deep learning model, Mask RCNN, for instance, segmentation to extract graphical elements from infographics. 
    The extracted graphical elements are used as visualization templates for creating similar infographics using different data.
    
    Meanwhile, since contents of bitmap visualizations are usually hard to be automatically extracted, some studies have proposed methods to embed the required information (\eg, meta data, color schema) into bitmap visualizations~\cite{zhang2020viscode, fu2020chartem}. The embedded information will not influence human perception of the visualization and can then be easily extracted from the visualization.
    For example, VisCode~\cite{zhang2020viscode} trains an encoder-decoder network, in which the encoder embeds a QR code to the background of a bitmap visualization and the decoder extracts the QR code from the visualization. A BASNet is trained to identify the semantically important regions and ensures the encoder will not affect human perception of these regions. 
    
    % Understanding the content enables the analysis of a large number of visualizations.
    % For example, Beagle~\cite{battle2018beagle} mines the web for a large corpus of visualizations and automatically classifies them by type (\eg,bar, pie). 
    % Based on the content understanding of these visualizations,
    % Beagle reveals the role of visualization in the real online world by answering questions like \textit{``What is the most popular visualization design?''}
    
    \noindent $\blacktriangleright$ \emph{Interpreting content.}
    More than extracting content, 
    % ML techniques can also be used to understand the content.
    ML techniques can also achieve a high-level interpretation of visualizations and solve complex tasks.
    % high-level understandings of visualizations to solve tasks that need complex reasoning,
    % and complex reasoning 
    For example, Kembhavi et al.~\cite{Kembhavi2016diagram} proposed an LSTM-based method to understand the concepts and relationships in a diagram.
    Questions about the diagram like \textit{``will the increases in lion lead to the decreases in deer?''} can be answered by an attention-based model with high accuracy.
    Another recent example DVQA is presented by Kafle et al.~\cite{kafle2018dvqa}. DVQA combines a CNN-based and an LSTM-based network and can answer questions about data visualizations.
    Given a visualization, the model reads the visualization content and answers reasoning questions such as
    \emph{``which item sold the most units?''}
    % These tasks usually involve multiple deep learning models.
% DVQA~\cite{} is an example,
% which attempts to parse barcharts 
% and answers the user's questions, \emph{e.g.}, \emph{are the values in the chart presented in a logarithmic scale?}

\noindent $\blacktriangleright$ \emph{Estimating human perception.}     
    % An ML model can provide useful suggestions during the design process by simulating human perception on visualizations.
    ML techniques can model how humans perceive a visualization and help designers better
    assess visualizations and understand audiences.
    A number of ML-based methods have been contributed to model user perception of a visualization from various aspects, including visual attention~\cite{bylinskii2017learning}, similarity recognition~\cite{ma2018scatternet}, and readability assessment~\cite{haleem2019evaluating}.
    For example, Bylinskii et al.~\cite{bylinskii2017learning} trained a deep learning model to predict the distribution of users' visual attention on an infographic.
    Guided by this prediction, designers can better arrange the layout and ensure that the most important contents are emphasized in the infographic.
    Haleem et al.~\cite{haleem2019evaluating} proposed a CNN-based approach to evaluate the readability of graph layouts by using graph images as inputs.
    An interesting phenomenon is that current ML4VIS studies extensively study the perception modeling in scatter plots, such as the perception of class separation~\cite{aupetit2016sepme}, the perception of local patterns~\cite{chegini2018interactive}, and the perception of similarity~\cite{ma2018scatternet}.
    This focus on scatter plots might stem from the wide popularity of scatter plots as well as their relatively simple visual forms.

\begin{figure}[]
  \centering
  \includegraphics[width=0.9\linewidth]{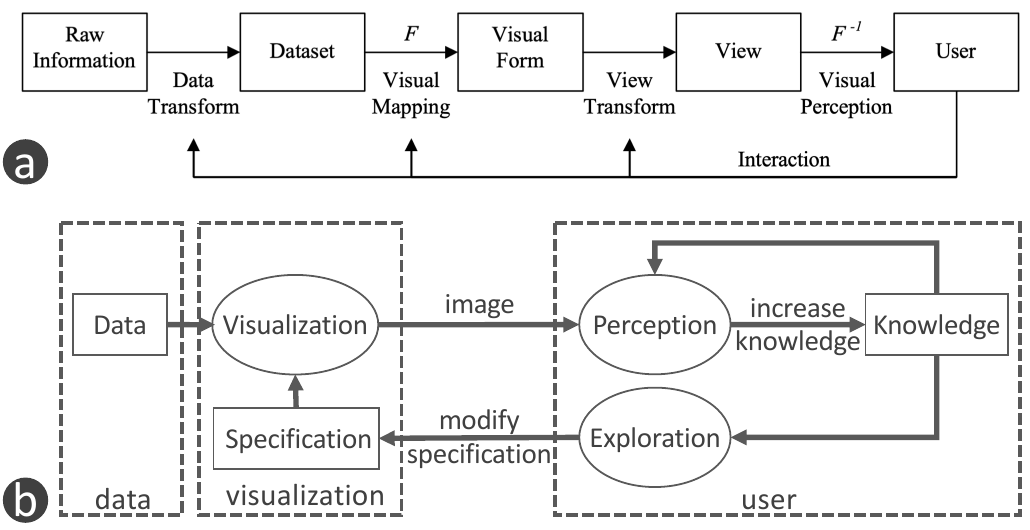}
  \caption{(a) A visualization pipeline proposed by Card~\cite{card1999readings}. (b) A model of visualization proposed by Van Wijk~\cite{van2005value}. }
  \label{fig:others_pipeline}
\end{figure}

\begin{figure}[]
  \centering
  \includegraphics[width=\linewidth]{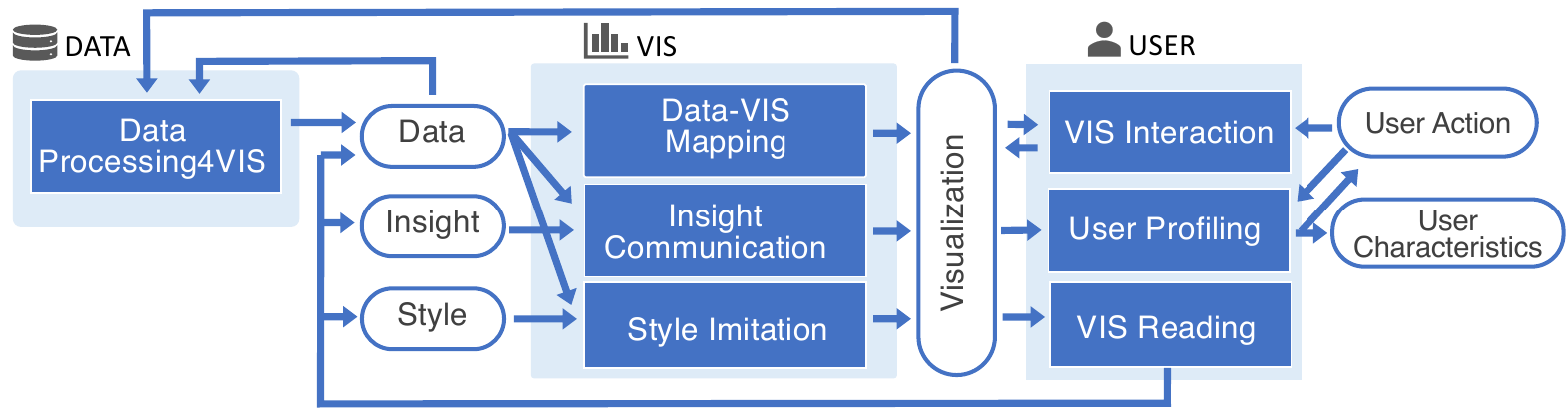}
  \caption{
  \revision{
  An ML4VIS pipeline connects the seven visualization processes that can be assisted by ML. Each blue rectangle presents a visualization process where ML can assist in. Each white circle indicates a key element and can be either the input or the output of a process. The arrows connecting the rectangles and circles indicate the input and output of each process. 
  }
  }
  \label{fig:ML4VIS_pipeline}
\end{figure}

\begin{figure}[]
  \centering
  \includegraphics[width=\linewidth]{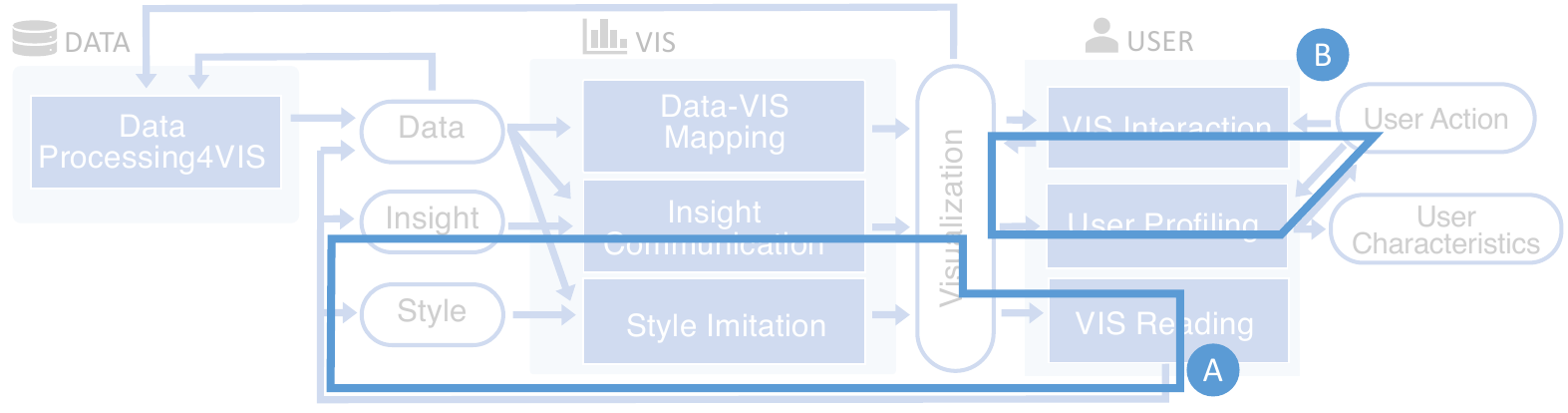}
  \caption{ 
  \revision{
  The employment of ML in one visualization process can benefit other processes and further improve the whole workflow.
  Analyzed examples: A) Bylinskii et al.~\cite{bylinskii2017learning} employed ML in \emph{VIS Reading} and further improved a manual \emph{Insight Communication} process. B) Ottley et al.~\cite{ottley2019follow} employed ML in \emph{User Profiling} and further improved an non-ML assisted \emph{VIS Interaction} process.
  }
  }
  \label{fig:pipeline_example}
\end{figure}

\section{An ML4VIS Pipeline}
\label{sec:ml4vis_pipeline}

To map out the role of ML4VIS in general visualization and better inform future exploration, we connect the seven processes to previous visualization models and present an ML4VIS pipeline.

This ML4VIS pipeline is inspired by existing visualization models, especially the one proposed by Card~\cite{card1999readings} and the one by Van Wijk~\cite{van2005value} (\autoref{fig:others_pipeline}).  
% We modify these visualization models to better fit the context of ML4VIS.
\revision{We modify these visualization models to better fit the scenarios of ML4VIS, \ie, to describe the visualization processes so that they can be better aligned with ML tasks.}
Specifically, we make the following main modifications.
First, contrary to previous studies, we take an additional element ``style'' into consideration. 
\revision{We refer to \emph{style} as elements that affect the appearance of a visualization but are not mapped to data values. Typical visualization styles include color palette, layout, and chart decoration.}
\emph{Style} is tightly related to the memorability~\cite{borkin2013makes} and engagement~\cite{segel2010narrative} of a visualization and \revision{has gained} increasing research attention.
Second, we decouple this pipeline from users' internal states, \eg, users' internal knowledge in Van Wijk's model, to better fit the context of ML4VIS.
Instead of including users' internal states, 
% In this pipeline, 
we use the outputs of these internal states (\eg, insights, user action) to ensure elements in the pipeline are describable instances rather than vague concepts. 
For example, user actions can reflect users' intentions and tasks and insights can reflect users' perceived knowledge from a visualization.
% which makes this pipeline better fit the context of ML4VIS as ML is good at well defined inputs and outputs.
\revision{Since ML usually requires well-defined inputs and outputs, the modified pipeline can better fit the context of ML4VIS by removing vague or hard-to-define concepts.
Meanwhile, as in Van Wijk’s pipeline, we also do not differentiate \textit{visual mapping} and \textit{view transformation} that are introduced in Card's pipeline. 
This decision is based on our observation of the collection of ML4VIS studies. 
In the collected ML4VIS studies, the visual mapping and the view transformation are sometimes achieved by an end-to-end ML model and are not independent (\eg, \cite{han2019tsr, chen2019generativemap}). 
}

% ensure that each process purely transfroms between describable elements.
% By puring mapping from elements to element, better fit the context of ML4VIS.
% 1. consider styles, increasingly attention in the field of visualization as the memorability, 
% 2. focus on initinaizable process, user interal state are not included

\renewcommand{\arraystretch}{1.2} %<- modify value to suit your needs
\begin{table*}[b]
    \centering
    \caption{Examples of Different Training Data Formats.}
    % \begin{tabular}{m{1.2cm}| m{0.6cm}| m{1.4cm} | m{0.8cm} | m{0.8cm} | m{0.8cm}}
    \begin{tabular}{m{1.6cm} *{6}{|m{2.3cm}}}
    \toprule
     & Data & Visualization & User Action & User Characteristic & Style & Insight \\ 
     \midrule
    Image & --- &\cite{poco2017extracting, poco2017reverse, jung2017chartsense, bylinskii2017understanding, kafle2018dvqa, chen2019towards, kahou2017figureqa, siegel2016figureseer, zhang2020viscode, chen2019generativemap, he2019insitunet} 
    & \cite{FanH2018fast, fan2019personalized} & ---
    & \cite{chen2019towards, lu2020exploring} &\cite{madan2018synthetically, bylinskii2017understanding, fosco2020predicting} \\
    \hline
    Natural \newline Language& --- & --- & \cite{yu2019flowsense, huang2019natural} & --- & --- & \cite{cui2019text, liu2020autocaption, jung2017chartsense, kafle2018dvqa, kim2018dynamic} \\
    \hline
    Engineered \newline Features & \cite{vizML2019CHI, luo2018deepeye}  & \cite{savva2011revision, battle2018beagle}  & \cite{steichen2013user, Lalle2015curve, gramazio2017analysis} & \cite{dereck2014, Lalle2015curve} & \cite{haleem2019evaluating} & --- \\
    \hline
    Sequences & \cite{data2vis2019CGA} & \cite{data2vis2019CGA, zhou2020table2charts} & \cite{brown2014finding, milo2018next, ottley2019follow, wall2019markov} & --- & --- & --- \\
    \hline
    Graph & \cite{wang2019deepdrawing, kwon2019deep} & --- & --- & --- & --- & \cite{Kembhavi2016diagram, kim2018dynamic} \\
    \bottomrule
    \end{tabular}
    
    \label{tab:data_format}
\end{table*}

The ML4VIS pipeline is presented in \autoref{fig:ML4VIS_pipeline}.
Each of the seven blue boxes indicates a process where ML can assist in; each of the white boxes indicates a key element that is either the input or the output of a process. 
As with Van Wijk's model~\cite{van2005value}, the seven processes are grouped by three containers: \emph{DATA}, \emph{VIS}, and \emph{USER}.

\noindent
\textbf{Example Usage:}
This ML4VIS pipeline illustrates the close relationship among the seven processes: the outputs of one process are used as the inputs for other processes.
Therefore, employing ML techniques in one visualization process can further facilitate other processes, even though ML techniques are not used in these processes.
When designing visualization tools, this ML4VIS pipeline can help decide where to embed ML techniques to enhance the whole workflow.
Below we present several examples in the collected papers as a demonstration.
\begin{itemize}[leftmargin=*]
 \item Bylinskii et al.~\cite{bylinskii2017learning} proposed a deep learning method to estimate the visual attention of audiences on different regions of an infographic.
The employed ML technique takes visualization as input, estimates user attention, and predicts user-perceived insights (\ie, information in the highlighted regions), thus contributing to \emph{User Profiling}.
Based on this method, the authors developed a tool to help designers evaluate their visualizations by checking whether important information received sufficient attention.
Meanwhile, the outputted insights can be used as input of \emph{Insight Communication}. In the developed authoring tool, designers can also interactively modify their infographic, observe the change of audience attention, and thus create visualizations that better communicate important insights.
As shown in \autoref{fig:pipeline_example}(A), the developed tool forms an \emph{Insight Communication}-\emph{VIS Reading} loop:
the \emph{Insight Communication} process does not employ ML techniques but benefits from the ML techniques used in \emph{VIS Reading}.
 \item Ottley et al~\cite{ottley2019follow} applied a Markov model to learn users' needs based on their mouse clicks and predicted users' next-step selections. 
 While the ML techniques were employed in \emph{User Profiling}, the predicted next-step selections can be used as input to the \emph{VIS Interaction} and help the visualization tool fetch potentially related data in advance, thereby improving the interaction latency in large-scale data analysis.
 As a result, the \emph{VIS Interaction} process, even though employing no ML techniques, can benefit from the ML techniques used in \emph{User Profiling}.
 As shown in \autoref{fig:pipeline_example}(B), \emph{VIS Interaction} and \emph{User Profiling} form a loop.
\end{itemize}
Meanwhile, the unconnected items in this pipeline can illustrate possible future research directions.
For example, the user characteristics extracted from \emph{User Profiling} can be used as inputs of \emph{Data-VIS Mapping}, \emph{Insight Communication}, and \emph{Style Imitation} to construct user-adaptive visualizations, \ie, visualizations that consider an individual user's needs, abilities, and preferences.
While several studies \cite{steichen2013user, dereck2014, Lalle2015curve} have discussed this direction in their future work, few ML techniques have been employed to successfully achieve this goal~\cite{gotz2009behavior}.

% This ML4VIS pipeline also illustrates the close relationship among the seven processes, which are not mutually exclusive and sometimes integrated with each other. 
% % We would like to emphasize that the seven processes are not mutually exclusive and sometimes integrated with each other.
% An example is the possible overlapping between \emph{\revision{VIS Reading}} and \emph{Insight Communication}. 
% Bylinskii et al.~\cite{bylinskii2017learning} contributed to the \emph{\revision{VIS Reading}} process by proposing a deep learning method to estimate users' visual attention on an infographic.
% At the same time, this deep learning method can also be used in \emph{Insight Communication} by guiding the layout of infographics to highlight important insights.\looseness=-1
% Another example is the 
% Gotz et al. \cite{gotz2009behavior} make behavior-driven visualization recommendations to facilitate \emph{\revision{Data-VIS Mapping}} based on the recognized patterns from \emph{VIS Interaction}.

% Another example is the overlapping between \emph{Style Imitation} and \emph{Insight Communication}.
% In DataShot~\cite{wang2019datashot}, a decision tree model learns from  trained a model to map from the user given insight to an infographic by learning from a collection of infographics. The model, while learning the map, inevitably learns the style of these infographics, such as the styles and the layout of graphic elements.

\section{Align ML with VIS}
\label{sec:map_ml}

In this section, we review the current ML4VIS studies from an ML perspective, aiming to answer the question \textit{``how is ML used to solve visualization problems''}. 
\revision{
We analyze both the training data and the ML tasks.
We first discuss the commonly-used formats of training data, which are summarized from the collected \paperNum{} ML4VIS studies using a bottom-up approach.
We then categorize the collected ML4VIS papers based on the main ML tasks \cite{nicolas2015scala, murphy2012machine} to provide an overview of how the needs in visualization are formed and solved as ML problems.
This ML perspective (\ie, training data, ML tasks) naturally corresponds to the visualization perspective (\ie, process inputs and outputs, visualization processes).
We map the summarized data formats to the six types of information (\ie, inputs and outputs of the visualization processes) and the ML tasks to the seven visualization processes.
}
This ML-VIS mapping aims to provide both an understanding of the current practices and a guidance for future exploration in the ML4VIS research.
% Unlike \autoref{sec:six_process} where the capabilities of ML are discussed with real examples, this section aims to offer a statistical analysis of the capabilities of ML in ML4VIS from the aspect of learning tasks.

\subsection{Training Data Formats}

% One strong advantage of ML4VIS studies is the ability to learn knowledge from the training dataset, which eliminates the dependency on human expertise.
The format of training data reflects how information is encoded in ML. It can influence how to formalize ML problems and choose suitable ML models.
The seven visualization processes in \autoref{sec:six_process} reveal six types of information: data (underlying the visualization), visualization, user action, user characteristics, style, and insight.
In this subsection, we summarize the common data formats that are used to encode the five types of information based on a bottom-up analysis of the collected papers (as shown in \autoref{tab:data_format}).
% The five types of information are encoded in different formats, as shown in \autoref{tab:data_format}.
% In this subsection, we summarize the commonly-used data formats in the collected \paperNum{} papers using a bottom-up approach and discuss how these data formats are used to encode the five types of information.

% We do not distinguish input or output data 
\textbf{Image} is the format that most visualizations ``in the wild'' are presented in~\cite{savva2011revision}. 
It is not surprising that many studies (\eg, \cite{poco2017extracting, poco2017reverse, jung2017chartsense, bylinskii2017understanding}) encode \textbf{visualizations} in the form of images, contributing a number of large datasets of real-world visualization images: 
FigureSeer Dataset (60k)~\cite{siegel2016figureseer}, AI2D dataset (5k) \cite{Kembhavi2016diagram}, Visually29K dataset (29k) \cite{madan2018synthetically}, DVQA dataset (300k) \cite{kafle2018dvqa}, FigureQA dataset (100k) \cite{kahou2017figureqa}, ColorMapping dataset (1.6k) \cite{poco2017extracting}.
The emergence of large visualization-image datasets may result from the considerable differences between natural images and visualization images.
Such differences make it difficult to directly apply the datasets and techniques developed for natural images to visualization images~\cite{Kembhavi2016diagram, chen2019towards, kim2018dynamic}. 
% Apart from visualization, \textbf{styles} and \textbf{insights} can be encoded as parts of an image.
% \yong{need further checking.}
The \textbf{styles} of visualizations and the \textbf{insights} extracted from visualizations can also be encoded as parts of an image.
For example, Chen et al. \cite{chen2019towards} encoded styles of a visualization as its segmented graphical icons.
Bylinskii et al. \cite{bylinskii2017understanding} encoded the insights of an infographic as visual elements that are diagnostic of the topic of an infographic.
% such as subregions~\cite{bylinskii2017learning} or segmented graphical elements~\cite{chen2019towards} of an image. 
Meanwhile, some researchers encode \textbf{user actions} in images.
For example, Fan et al.~\cite{FanH2018fast} encoded the user actions (\ie, click and drag) in a scatter plot as the zoom and rotation of the scatter plot image.\looseness=-1

\textbf{Natural language} (\eg, text and speech) is the most common form of human communication~\cite{chowdhary2020natural}.
Prior studies~\cite{yu2019flowsense, huang2019natural} have enabled users to use natural language to express intended \textbf{user actions}, aiming to achieve more convenient and intuitive \emph{VIS Interaction}.
For example, Huang et al.~\cite{huang2019natural} allowed users to interact with a visualization system
using textual sentences such as ``visualize trajectories passed through tourist attractions during January 25''.
Apart from user actions, \textbf{insights} can also be encoded in the form of natural language.
Through natural language, users can easily describe the insights they aim to convey in \emph{Insight Communication}~\cite{cui2019text}
and understand the insights extracted in \emph{\revision{VIS Reading}}~\cite{kafle2018dvqa, Kembhavi2016diagram, liu2020autocaption}.
For example, Text-to-Viz~\cite{cui2019text} allows users to describe the insights using a natural language statement and automatically generates the corresponding infographics.  
AutoCaption~\cite{liu2020autocaption} generates a text caption to describe the insights of a visualization based on four types of chart features detected (\ie, aggregation, comparison, trend, distribution).
In contrast to images,
researchers usually collected a small number of labeled natural language examples for their specific tasks.
For example,
% FlowSense
Yu and Silva~\cite{yu2019flowsense} collected and labeled less than 20 examples to train an ML model that can resolve certain syntactic ambiguity;
% Text-to-Viz
Cui et al.~\cite{cui2019text}
collected and labeled 800 examples to train an ML model that can extract the four predefined entities.
% The small number of training data might result from the combination of ML-based and non-ML-based methods.
Also, due to the small number of training data, ML-based methods are often combined with non-ML-based methods.
For example, Yu and Silva~\cite{yu2019flowsense} supported natural language queries using non-ML-based semantic parsing. An ML model is only trained on few labeled examples to resolve certain syntactic ambiguities.

% The small number of labeled examples might result from the similarity between the natural language in ML4VIS and the general natural language.
% The datasets and algorithms developed for general natural language tasks can be easily applied to ML4VIS.
% For example, Huang et al.~\cite{huang2019natural} used the Word2Vec algorithm and the Wikipedia dataset to find similar words for a query word in a natural-language-based visual query
% of uncertain human trajectories.

\textbf{Engineered features} are high dimensional vectors that are extracted from the raw data using domain knowledge.
While the great progress of deep learning helps reduce the reliance on human-centric feature engineering, extracting features using human knowledge is still popular in ML4VIS studies.
This popularity of feature engineering may come from its ability to effectively capture important information without the need for large datasets and enormous computational resources.
Human specified features have been used to encode all the five types of information except insights.
% , as the examples blow show.
Here we discuss several representative examples.
For the \textbf{data} (underlying a visualization), VizML~\cite{vizML2019CHI} encodes a tabular data using 841 features related to statistical analysis, such as mean, standard deviation, entropy, and skewness.
For the \textbf{visualization}, Beagle~\cite{battle2018beagle} encodes visualizations using the features extracted from SVG visualization files, such as the number of axes, the position of circles, and the width of rectangles. 
For the \textbf{user actions}, Gramazio et al.~\cite{gramazio2017analysis} encoded user actions (\ie, mouse clicks) through a set of human-defined features, such as the dwell time, the active time, the region-of-interest transitions.
For the \textbf{style}, Haleem et al.~\cite{haleem2019evaluating} summarized the style of a graph layout using a set of graph aesthetic metrics, such as the node occlusion and the edge crossing.

\textbf{Sequences} are rows of values whose order is important. 
Sequences have been widely used to encode \textbf{user actions} in their temporal order~\cite{ottley2019follow, brown2014finding, wall2019markov}.
For example, Brown et al.~\cite{brown2014finding} conducted a sequence-based analysis of the user actions in visual analytics.
The sequences of seven types of user actions (\ie, pan left, right, up, down, zoom in, and out) are converted into sub-sequences of two or three user actions, such as first zoom in and then zoom out.
% Sequences are also used to encode \textbf{data} (underlying visualizations) and \textbf{visualizations}.
Interestingly,
even though \textbf{data} (underlying visualizations) and \textbf{visualizations} do not have a sequential structure, they can also be encoded using sequences.
For example, Data2Vis~\cite{data2vis2019CGA} encodes both JSON data file and Vega-lite visualization configuration as sequences.
Table2Chart~\cite{zhou2020table2charts} encodes a visualization as a sequence of actions, including selecting data fields, selecting chart types, splitting data fields, and grouping data fields. \looseness =-1

% \textbf{Table} is a widely used format to present the data underlying visualizations~\cite{viznet}.
% Data (underlying visualizations) can be directly encoded in its original table format in ML4VIS studies~\cite{zhou2020table2charts, }.
% Meanwhile, insights can also be represented as part of a table, \ie, columns of the table that have interesting patterns~\cite{demiralp2017foresight}.
% \\

% \textbf{Multivariate data} is a widely used format to present the data underlying visualizations~\cite{viznet}.
% Encoded using this format, data (underlying visualizations) can be used in their original formats (\eg, table, 3D shapes) and achieve end-to-end learning.

\textbf{Graph} is a non-Euclidean data format that consists of nodes and edges.
\textbf{Data} (underlying visualizations) can be encoded in its original graph format in ML4VIS studies~\cite{wang2019deepdrawing, kwon2019deep}.
% can also be used to organize user actions in a semantic way.
Meanwhile, \textbf{insights} of visualizations can also be encoded in the form of graphs~\cite{Kembhavi2016diagram, kim2018dynamic}.
For example, Kim et al.~\cite{kim2018dynamic} encoded the insight of an infographic as a graph. 
Each node of the graph represents an object in the infographic and each edge represents the relationship between the two connected objects.

\subsection{Mapping ML Tasks to Visualization Processes}

This subsection first provides a statistic summary of how the different ML tasks are mapped into different visualization processes.
Then, we discuss each learning task in detail, including the related visualization processes, the commonly-used ML models, and the representative examples.

When summarizing the collected studies from an ML perspective, we find that it is not practical to summarize based on the employed ML models,
as new ML models are constantly emerging and one ML model is often applied to solve different problems.
To better align the capabilities of ML with the needs in visualization, we categorize the existing ML4VIS studies based on the types of ML tasks. 
% rather than the types of ML algorithms (\eg, Markov machine, convolutional neural network). 
% in order to align needs with capability
% have an overview of the capabilities of current ML methods and ,
This learning-task-based categorization allows us to understand how the needs in visualizations are formed and solved as ML problems, without diving into the ML model details (\eg, whether to use LSTM or GRU).
% In practices, one ML algorithm can be used for different learning tasks, \eg, CNN can be used for in regression, classification, and generative learning.

% Based on the types of learning, current ML methods can be categorized into four basic paradigms, \ie, supervised learning, unsupervised learning, semi-supervised learning, reinforcement learning~\cite{murphy2012machine, nicolas2015scala}. 
% We follow the taxonomy in \cite{nicolas2015scala} and further categorize the four paradigms into seven minimal subsets: clustering, dimension reduction, generative, classification, regression, semi-supervised learning, reinforcement learning, and semi-supervised learning, as shown in \autoref{tab:learning_task}.
% One minimal subset of learning tasks is obtained when the following categorization is no longer related to learning tasks.
% For example, Nicolas~\cite{nicolas2015scala} further categorized reinforcement learning as Markovian or evolutionary. Since this categorization is about learning mechanisms rather than learning tasks, we treat reinforcement learning as one minimal subset of learning tasks. 
Current ML methods can be categorized into four basic paradigms, \ie, supervised learning, unsupervised learning, semi-supervised learning, and reinforcement learning~\cite{murphy2012machine, nicolas2015scala}. 
Based on the collected \paperNum{} papers, we refer to the learning tasks discussed in \cite{murphy2012machine, nicolas2015scala} and further categorize the four paradigms into seven subsets: clustering, dimension reduction, generative, classification, regression, semi-supervised learning, and reinforcement learning.
One subset of learning tasks is obtained when the following categorization is no longer related to learning tasks.
For example, Nicolas~\cite{nicolas2015scala} further categorizes reinforcement learning as Markovian or evolutionary. Since this categorization is about learning mechanisms rather than learning tasks, we treat reinforcement learning as one minimal subset of learning tasks.

% Note that there is no standard learning task taxonomy. 
% % Several inconsistencies exist among different taxonomies.
% For example, while some taxonomies~\cite{nicolas2015scala} list semi-supervised learning as an independent learning diagram, others~\cite{tour2019, azure_algorithm} treat semi-supervised learning as a type of generalized supervised learning.
% After all, these differences among ML taxonomies are beyond the scope of this paper.

\begin{figure}[]
    \centering
    \includegraphics[width=\linewidth]{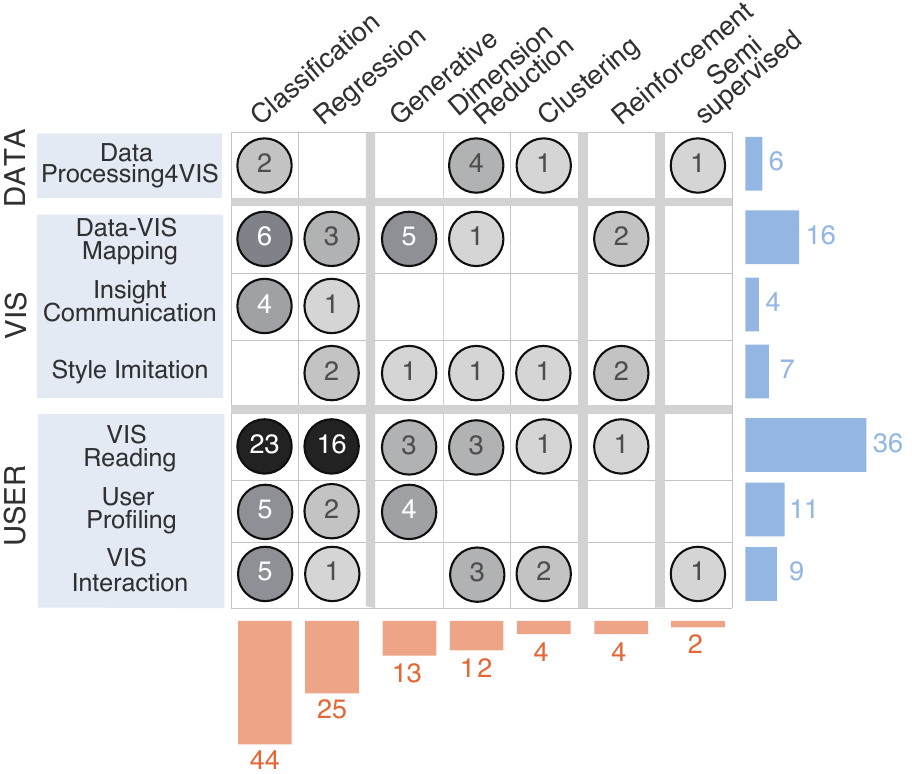}
    \caption{Each row is a visualization process and each column is an ML task. In each cell, the number and the grayscale indicate the number of papers.
    The height of the bar indicates the number of studies related to a type of ML task (blue bar) or a visualization process (orange bar).
    Note that the number on a bar can be smaller than the sum of the numbers in the corresponding row/column, because some papers involve multiple learning tasks or visualization processes. }
    \label{fig:map_ML4VIS}
\end{figure}

% \subsection{ML-VIS Mapping}
% \subsection{Current Practices in ML4VIS}
\subsubsection{An Overview}
\autoref{fig:map_ML4VIS} shows the ML-VIS mapping between the seven visualization processes and the ML tasks.
Each column represents one type of learning task and each row represents one visualization process.
\revision{
The columns are grouped using gray thick lines based on the four paradigms, \ie, supervised learning, unsupervised learning, semi-supervised learning, and reinforcement learning.
The four-column groups, and columns inside each group, are in descending order of the number of papers.
}
In each cell, the number and the grayscale indicate the number of studies that are related to both the ML task in column and the visualizations process in row.
The height of the bar indicates the number of studies related to a type of ML task (\textcolor{mlOrange}{orange} bar) or a visualization process (\textcolor{visBlue}{blue} bar).
Note that the number on a bar can be smaller than the sum of the numbers in corresponding cells, since a paper can involve more than one visualization process and one ML task.

%  within the corresponding learning task and visualization process.
% This ML-VIS mapping reflects how the needs in visualization are formed as ML learning problems in current ML4VIS studies.

Overall, most ML4VIS studies formulate the needs in visualization processes as supervised learning tasks (44 as classification and 25 as regression).
% Reinforcement learning, on the contrary, is least used in ML4VIS studies.
% In supervised learning,
% more studies formulate the problem as classification than regression problems (48 vs. 27).
% 34 (47\%) ML4VIS studies, covering five of the seven visualization processes, formulate the needs in visualization as (classification) problems.
Other types of learning tasks are less used.
Unsupervised learning, the second most-used learning, is employed in 27/\paperNum{} ML4VIS studies.
Only 4 papers use reinforcement learning and 2 papers use semi-supervised learning.
This is not surprising as supervised learning is the most common type of learning~\cite{murphy2012machine}.
When taking different visualization processes into consideration,
the preference towards supervised learning, especially classification, appears on most visualization processes.
\revision{
This preference is most pronounced in \emph{VIS Reading}, where 23/36 studies are related to classification.
Unlike other visualization processes,
in \emph{\revision{Data Processing4VIS}} and \emph{Style Imitation}, the collected ML4VIS studies exhibit a preference for unsupervised learning.
But this observation needs to be treated cautiously due to the small number of studies in the two processes.
}

Among the seven visualization processes, \emph{\revision{VIS Reading}} is the most frequently investigated visualization process in ML4VIS (by 36/\paperNum{} papers). 
Most of the studies in \emph{\revision{VIS Reading}} employed supervised learning (23/36 used classification and 16/36 used regression).
We speculate this is caused by the success of deep learning in computer vision tasks (\eg, image classification, object detection, instance segmentation).
A large number of deep learning models have been proposed to understand natural images and are adopted for the perception of visualization images~\cite{lai2020automatic, lu2020exploring, chen2019towards, bylinskii2017learning, madan2018synthetically}.
% Most studies in \emph{\revision{VIS Reading}} formulate as supervised learning 
% A considerable proportion of ML4VIS studies investigate the problems in
% \emph{VIS Interaction} (by 21 papers) and \emph{\revision{Data-VIS Mapping}} (by 16 papers).
\revision{
A considerable proportion of ML4VIS studies investigate the problems in \emph{\revision{Data-VIS Mapping}} (16 papers), \emph{User Profiling} (11 papers), and \emph{VIS Interaction} (9 papers).}
The other three processes (\ie, \emph{\revision{Data Processing4VIS}}, \emph{Insight Communication}, \emph{Style Imitation}) are relatively less investigated.

\subsubsection{Supervised Learning}
In supervised learning, a model learns the mapping from input $X$ to output $Y$ from the labeled training examples.
Labeled input-output pairs are required in supervised learning.
Supervised learning can be further divided into \textbf{regression} and \textbf{classification} based on whether the output is numerical or categorical~\cite{murphy2012machine}.

When formalizing a visualization problem as a \textbf{regression} task, the training examples need to be labeled with numerical values.
% Continuous values have been widely used to describe all the five types of information along the ML4VIS pipeline,
% such as the aesthetic scores of visualizations (style), the positions of graphical elements (visualization), the dwell time of a user click (user action), the user attention on different regions of a visualization (insight).
\revision{
Among all visualization processes,
regression is mainly (19/25) used in the USER stage, especially in \emph{VIS Reading} (16 papers).
For example, in \emph{\revision{VIS Reading}}, regression can be used to learn the salient regions of a visualization~\cite{bylinskii2017learning, madan2018synthetically}, the segmentation of important visual objects~\cite{chen2019towards}, and the aesthetic scores of visualizations~\cite{what2018kwon, haleem2019evaluating}.
In \emph{User Profiling}, regression can also be used to learn the dwell time of user mouse clicks~\cite{gramazio2017analysis} or the users' learning curve of understanding a visualization~\cite{Lalle2015curve}. 
Regression is not used in DATA stage and less used in the VIS stage, which may be due to the difficulty to quantitatively label the data processing and the visualization creation.
}
% When used in the VIS stage, regression is often used to accomplish a

% Many ML models can be used to generate continuous outputs. 
Many ML models can be used for regression tasks.
How to choose a suitable ML model largely depends on the data formats.
For example, CNNs are widely used when the input data are images.
Many CNN models that are originally proposed for general computer vision tasks have been adopted to process visualization images, including ResNet (used in~\cite{siegel2016figureseer}), Xception (used in~\cite{fosco2020predicting}), FCN used in (~\cite{Liu2018learning}), Fast R-CNN (used in~\cite{madan2018synthetically}), and Mask R-CNN (used in~\cite{chen2019towards, lai2020automatic}).
LSTM and GRU are used when processing sequences and natural languages~\cite{kafle2018dvqa, kim2018dynamic}. 
Traditional (\ie, non-deep-learning-based) ML models (\eg, SVR, logistic regression, MLP) are usually used when the inputs are engineered features.

When formalizing a visualization problem as a \textbf{classification} task, all the training examples need to be categorized into finite classes.
Classification has been widely used in ML4VIS studies, such as evaluating whether a visual encoding is valid~\cite{luo2018deepeye}, predicting whether a visual region has interesting patterns~\cite{lekschas2020peax}, and recognizing the types of charts~\cite{savva2011revision, jung2017chartsense}. 
Complicated problems can be decomposed into a series of classification tasks. 
For example, 
VizML~\cite{vizML2019CHI} decomposes the creation of a visualization as a series of classification tasks, including classifying the type of marks, the type of shared axis, and the type of visualization.
Kim et al.~\cite{kim2018dynamic} and Kembhavi et al.~\cite{Kembhavi2016diagram} decompose the understanding of a diagram infographic as classifying the relationships between every two visual objects in the diagram. 
Classification has been used in all the seven visualization processes except for \emph{Style Imitation}.
We conjecture this is because that the possible styles of a visualization can hardly be categorized into finite classes.

Most models used for regression can also be used for classification tasks through different ways, for example, 
setting a threshold (binary classification, 1-vs-the-rest classification) or using a softmax function (multi-class classification).
Meanwhile, there are some ML models whose outputs are, by nature, discrete, such as SVM and decision tree.

The recent success of end-to-end deep learning methods has contributed many ML models that can accomplish regression and classification tasks at the same time.
For example, Faster RCNN~\cite{ren2015faster}, an instance segmentation model, has been used in \revision{VIS Reading} and can predict the type (classification) and the bounding box (regression) of visual objects simultaneously~\cite{madan2018synthetically}.

% generative:
% VAE, Markov model, 

\subsubsection{Unsupervised Learning}

In unsupervised learning, a model learns the underlying structure of the unlabeled data $X$.
Compared with supervised learning, unsupervised learning relieves the requirement of the expensive data labeling process.
Based on the collected papers, we discuss the three common tasks when using unsupervised learning in ML4VIS: generative learning, dimension reduction, and clustering.

% generative
% define
Similar to prior research~\cite{kwon2019deep}, this paper refers a \textbf{generative} model to a model that learns the distribution of unlabeled data and
is capable of generating new samples that are similar to, but not the same as, the training data. 
% To be solved using a generative model, a problem needs to be formalized as finding a sample similar to the training data.
% For example,
% Generative is mainly used in four visualization processes: \emph{\revision{Data-VIS Mapping}}, \emph{Style Imitation}, \emph{\revision{VIS Reading}}, and \emph{VIS Interaction}.
\revision{Generative learning is mainly used in four visualization processes: \emph{Data-VIS Mapping}, \emph{Style Imitation}, \emph{VIS Reading}, and \emph{User Profiling}.}
In \emph{\revision{Data-VIS Mapping}} and \emph{Style Imitation}, the creation of a visualization can be formed as generating a visualization that is similar to the collected visualization examples~\cite{generativemodel2015tvcg, chen2019generativemap, he2019insitunet, kwon2019deep}.
\revision{
In \emph{User Profiling}, predicting user actions can be formed as generating the next user actions, so that the action sequence is similar to the collected user action sequences~\cite{ottley2019follow, milo2018next}.
}
In \emph{\revision{VIS Reading}}, the human perception of a visualization can be modeled through latent features~\cite{abbas2019clustme}.

% model & examples
Among all the generative models,
GAN is a hot research topic in the field of deep learning~\cite{gui2020review}.
In ML4VIS, GANs have been exclusively used in \emph{\revision{Data-VIS Mapping}} for scientific visualization, 
including volume rendering~\cite{generativemodel2015tvcg}, density maps~\cite{chen2019generativemap}, and in-situ visualization of ensemble simulations~\cite{he2019insitunet}.
GANs can effectively synthesize visualizations without actually running the rendering process, which can be time-consuming and computationally expensive. 
However, training GANs is usually difficult due to non-convergence and model collapse~\cite{goodfellow2016nips}. 
Therefore, 
traditional (non-deep-learning-based) generative models, such as Markov model, Gaussian mixture model, variational autoencoder, are also popular, especially when the training data is small and the problem is relatively simple.
For example, a click event has a relatively simple data format compared with a visualization image and thus can be generated using a less complicated ML model.
Ottley et al.~\cite{ottley2019follow} used a hidden Markov model to generate future click events on a scatter plot visualization. 
The evolution of user attention on the visualization is modeled based on the collected clickstream data.
The future click events are then generated based on the user attention evolution.

In \textbf{dimension reduction}, an ML model transforms the data from a high-dimensional space to a low-dimensional space.
% To be solved through dimension reduction, a problem needs to be formed as finding a low-dimensional representation of the original data. 
In the collected ML4VIS studies, dimension reduction is used in five visualization processes for two main purposes.
In \emph{\revision{Data Processing4VIS}} and \emph{\revision{Data-VIS Mapping}}, dimension reduction is often used to
process high-dimensional data for the visualization in a 2D space~\cite{wang2017perception, saha2017see}.
For example, to visualize streaming data with varying dimensions, Fujiwara et al.~\cite{fujiwara2019incremental} improved the incremental principal component analysis (PCA) and proposed a new dimension reduction solution.
In \emph{Style Imitation}, \emph{\revision{VIS Reading}}, and \emph{VIS Interaction},
dimension reduction is often used to extract representative features for effective analysis~\cite{jo2019disentangled, fujiwara2019incremental}. 
For example,
to characterize the distribution patterns showed in scatter plots, Jo and Seo~\cite{jo2019disentangled} trained a $\beta$-variational autoencoder to transform
a scatter plot image ($64 \times 64 \times 1$) to a $32 \times 1$ feature vector.
The 32 features capture the underlying data distribution of the scatter plot and can be used to predict the human-perceived distances between scatter plots. \looseness=-1

Popular dimension reduction methods include PCA, MDS, t-SNE, UMAP, and autoencoder.
PCA, MDS, t-SNE, and UMAP have been widely used for visualizing high dimensional data on the 2D space.
While the liner methods such as PCA and MDS are good at preserving global structures, non-linear methods such as t-SNE and UMAP are good at preserving local structures.
Autoencoders can be easily combined with other ML models to extract features from complicated inputs.
For example, an autoencoder can be combined with graph neural networks (GNNs) to extract features from graphs~\cite{kwon2019deep}, combined with 3D convolution to extract features from 3D streamlines and surfaces~\cite{han2020flownet}, and combined with CNNs to extract features from images~\cite{jo2019disentangled, fu2019vis_assessment}.

In \textbf{clustering}, a model divides unlabeled data into a number of groups based on their similarity.
% Based on the collected papers, clustering is mainly used in three visualization processes: \emph{Style Imitation}, \emph{\revision{VIS Reading}}, and \emph{VIS Interaction}. 
Clustering is a classical problem and has been used for many applications in the visualization field.
% We only examine studies where clustering is used to facilitate the creation, interaction, and evaluation of visualizations and exclude studies where clustering is purely used for the data analysis.
In \emph{Style Imitation}, collected training examples can be clustered based on their color styles~\cite{colorCrafting2020TVCG} to guide the creation of a new visualization. 
In \emph{VIS Interaction} and \emph{\revision{VIS Reading}},
visual objects in a visualization can be clustered to better understand and interact with this visualization~\cite{abbas2019clustme, fan2019KDE, han2020flownet}.
Generally speaking, clustering algorithms can achieve the most satisfactory results when the input features are representative.
Therefore, when clustering complicated inputs (\eg, images, natural language, extremely high-dimensional features), it is necessary to apply feature engineering or dimension reduction to extract representative features from the original data.
We refer the readers to \cite{murphy2012machine, gan2020dataclustering} for more details about clustering algorithms.

\subsubsection{Semi-supervised Learning}

In semi-supervised learning.
a model is trained using a small amount of labeled data with a large amount of unlabeled data.
Semi-supervised learning is a combination between unsupervised learning (no data are labeled) and supervised learning (all data are labeled).
Overall, semi-supervised learning is a less explored type of learning in ML4VIS.
The only two studies~\cite{lekschas2020peax, luo2020interactive} that employed semi-supervised learning belong to the same sub-category, \ie, active learning.
More specifically, Luo et al.~\cite{luo2020interactive} employed active learning to interactively clean data in \emph{\revision{Data Processing4VIS}}. Users are asked to label whether certain data items are duplicated and will influence the visualization.
Their labels will then be used to improve the classification of data duplicates.
Lekschas et al.~\cite{lekschas2020peax} applied active learning to improve the interactive visual pattern search in \emph{VIS Interaction}.
A set of unlabeled visual regions are identified using an active learning strategy, and labeled by the users based on whether they contain interesting visual patterns. 
A classification model is then iteratively trained using the user labels to predict the interestingness of other unlabeled regions of the visualization.

\subsubsection{Reinforcement Learning}

In reinforcement learning, an agent learns to take actions in an environment to maximize the cumulative rewards.
% Unlike supervised learning, which learns from the labeled data,
% reinforcement learning learns from action rewards in a trial-and-error manner.
Different from supervised and unsupervised learning, the task of reinforcement learning is not to directly generate end results (\eg, classification labels, latent features) but to take actions in an interactive environment (\eg, playing a GO game).

To be solved through reinforcement learning, a problem needs to be formalized as finding a sequence of actions that maximize the cumulative rewards.
More specifically,
the problem needs to be described as a Markov decision process, which contains a state space $S$, an action space $A$, a transition function, and a reward function.
At each time step, the agent is in a certain state $s_t \in S$ and chooses an action $a_u \in A$. 
The agent moves to a new state ${s_{t+1}}$ in the next time step based on the transition function $s_{t+1}=T(s_t, a_u)$ and gets a corresponding reward $r$ based on the reward function $s_{t+1}=T(s_t, a_u)$ \cite{sutton2018reinforcement}.
% Therefore, reinforcement learning can help solve more complicated problems that is beyond a simple mapping from inputs to outputs.
For example, PlotThread~\cite{tang2020plotthread} formalizes the problem of creating a user-preferred storyline layout as performing a sequence of layout modifications on a less satisfying layout. 
All possible layouts form the state space and all possible layout modifications form the action space.
The reward is the similarity between the current layout and the user-preferred layout.
The transition is implemented by applying the layout modification to the current layout.

Reinforcement learning has made significant success in a variety of tasks and a large number of reinforcement learning models have been proposed.
However, only a limited number of ML4VIS studies have used reinforcement learning, including asynchronous advantage actor-critic~\cite{mnih2016A3C} (used in PlotThread~\cite{tang2020plotthread}),
policy gradient~\cite{sutton2000policy} (used in MobileVisFixer~\cite{wu2020mobilevisfixer}),
and deep Q-learning~\cite{mnih2015DQN} (used in Table2Chart~\cite{zhou2020table2charts}).
How to apply reinforcement learning to ML4VIS still requires further exploration.
% Double Thompson Sampling for Dueling Bandits

% The two ML4VIS processes as the USER stages a large propotion of the collected ML4VIS studies (59, 59.7\%).

\section{Research Challenges \& Opportunities}
\label{sec:opportunities}

In this section, we discuss the research challenges and opportunities of ML4VIS that are derived from our survey, \qianwen{hoping to provide insights for both researchers and practitioners in visualization}. 
% \qianwen{We hope these discussions can provide insights for both researchers and practitioners and promote the exploration of ML4VIS.}

\subsection{From the Aspect of ML}
\textbf{More Diverse Types of ML:}
% \textbf{Limited Types of ML}
According to our survey, semi-supervised learning and reinforcement learning are less considered in ML4VIS studies, even though they demonstrate promising properties for ML4VIS studies.\looseness=-1

Semi-supervised learning extracts knowledge from data that only a small proportion is labeled. 
By combining unlabeled data with labeled data, semi-supervised learning not only reduces the expense of data labeling but also surpasses both supervised learning (trained only using the \qianwen{small amount of} labeled data) and unsupervised learning (trained using all data without their labels).
% Data labeling can be expensive in visualizations, especially when the labeling requires expertise
% As a result, semi-supervised learning can reduce the high expense of data labeling, .
For many problems in ML4VIS (\eg, generating suitable visualizations), the data labeling requires human experts (\eg, skilled designers).
Sometimes, it is infeasible to get a large and fully labeled training dataset.
The application of semi-supervised learning can help address this issue by reducing the required number of labels.
Initial exploration of semi-supervised learning in ML4VIS has been conducted.
For example, Lekschas et al.~\cite{lekschas2020peax} and Luo et al.~\cite{luo2020interactive} have demonstrated the effectiveness of using active learning in data cleaning and visual pattern search. 
However, how to use more techniques in semi-supervised learning for more visualization processes in ML4VIS still requires further investigation.

Reinforcement learning is able to learn sequential decisions without enumerating all possible training examples.
\qianwen{
While current ML4VIS studies that use reinforcement learning are relatively sparse (4/\paperNum{}),
previous ML studies have successfully applied reinforcement learning to human-computer interaction and demonstrated the capabilities of reinforcement learning in modeling user-data interactions~\cite{McCamish2018data, grotov2016online} and in making personalized recommendations~\cite{zheng2018drn}.
These previous studies in general user interaction show the potential of reinforcement learning in various visualization problems,
such as predicting next-step interaction in \emph{VIS Interaction}, modeling users in \emph{\revision{User Profiling}}, and recommending suitable visual representations in \emph{\revision{Data-VIS Mapping}} and \emph{Insight Communication}.}

\qianwen{Note that we summarize the collected ML4VIS studies in terms of ML tasks to better align ML capabilities with visualization needs in this survey. We also encourage the readers to consider ML from other perspectives and embrace more diverse ML techniques (\eg, federated learning, transform learning).}

\vspace{0.5em}\noindent
\textbf{Public High-quality Datasets:}
 In this survey, we notice that most papers need to construct their own datasets due to the lack of public visualization datasets~\cite{viznet, luo2018deepeye, vizML2019CHI}. 
%  This phenomenon reflects the lack of public datasets and benchmark tasks in ML4VIS.
%  More importantly, the quality of datasets need to be carefully assessed so that it will not endanger the validity of the obtained ML models. 
In existing ML4VIS studies,
the dataset quality is often limited by the size of data and the reliability of the label, and may potentially endanger the validity of the obtained ML models. 
 For example, DeepEye\cite{luo2018deepeye} learns to classify ``good''/``bad'' visualizations based on the training examples labeled by 100 students, whose knowledge about visualization is unclear.
 VizML~\cite{vizML2019CHI} trains a visualization recommender using collected online visualizations, yet previous studies have pointed out that online visualizations have a large proportion of deceptive visualizations~\cite{correll2017black}.
%  yet the quality of online visualizations has been challenged in previous studies~\cite{correll2017black}.
%  VizML~\cite{vizML2019CHI} assumes the collected online visualizations are made with professional design choices and can teach an ML model to recommend visualization design choices, even though previous studies have pointed out the biquity of low-quality, even deceptive, visualizations online~\cite{correll2017black}.
These studies prove the effectiveness of ML in solving visualization problems, but the questionable quality of their training data can degrade the performances of the ML models (\eg, the quality of the recommended visualizations).
Public and high-quality datasets are needed to further improve the application of ML in visualizations.
We hope this survey can encourage more research on related directions.

\vspace{0.5em}\noindent
\textbf{Benchmark Tasks:}
% \textbf{Lack of Benchmark Tasks}
ML4VIS studies are still at their early stage and benchmark tasks for ML4VIS remain unclear. 
However,
benchmark tasks are important for the progress of ML techniques~\cite{murphy2012machine}.
In the field of ML, 
ML models are commonly evaluated by comparing their performances with the state-of-the-art on benchmark tasks (\eg, question answering, instance segmentation).
Advanced tasks can be accomplished by combining the methods developed for multiple benchmark tasks.
% Performances evaluation is usually conducted by
The lack of benchmark tasks in ML4VIS makes peer comparison difficult and can lead to negative impacts on follow-up studies.

Meanwhile, we find that existing ML4VIS studies have intensively investigated several tasks, including 
graphic elements extraction~\cite{chen2019towards,poco2017reverse},
visualizations generation~\cite{data2vis2019CGA,vizML2019CHI,luo2018deepeye},
and graph layout~\cite{kwon2019deep, wang2019deepdrawing,what2018kwon}.
These tasks can be the start point to define benchmark tasks in ML4VIS.

\subsection{From the Aspect of VIS}
% 1. some topics are more hot than others.
% for the type of graphs, mainly focus on graph, scatter plot, bar charts.
% for the tasks, many color extraction.
% More advanced tasks: a multi-view VA?
% still a long way to go
\vspace{0.5em}\noindent
\textbf{Diversity in Visualization}
We notice that the forms of visualization are limited in the collected papers. 
First, current ML4VIS studies exclusively focus on static visualizations. 
Even though other forms of visualizations, such as animated transitions and data videos, are popular and effective in information communication, these visualizations are rarely discussed in the collected ML4VIS studies.
We conjecture this is partly caused by the difficulty of understanding and generating dynamic visualizations using ML techniques.
Take image-format training data as an example.
Compared with statistic visualization (images),
dynamic visualizations (videos) require a larger set of training data, more complicated ML models, and more powerful computational resources.
Second, most ML4VIS studies focus on certain types of standard visualizations, especially bar charts, scatter plots, and graphs. 
Other types of visualizations, such as treemap, streamgraph, and parallel coordinates, are not discussed despite their wide popularity in the real world.
Moreover,
chart decorations are overlooked even though they are important for the memorability and user engagement of visualization. Multi-view visualizations are rarely mentioned even though they are commonly used in visual analysis.
Since the success of visualization depends on choosing proper forms of visualization~\cite{munzner2014visualization},
we believe good opportunities exist in applying ML techniques to more diverse visualizations.

\textbf{Towards VIS-Tailored ML:}
In this survey, most ML4VIS studies directly apply ML techniques developed by ML researchers.
However, general ML techniques do not always suit well for the specific problems in visualization.
Take computer vision in \emph{\revision{VIS Reading}} as an example. 
General ML techniques are mainly developed for natural images and cannot be directly applied to images with charts and diagrams~\cite{Kembhavi2016diagram,chen2019towards}.
Chen et al.\cite{chen2019towards} employed a deep learning model to understand timeline infographics, but found that they needed a series of post-processing to address the poor model performance caused by the difference between natural images and charts.
The authors also highlighted the importance to develop ML techniques specified for visualizations.
% Considerable modifications, even novel structures, of ML  are needed in ML4VIS.
% We hope to see more studies that propose novel ML techniques to better solve the visualization problems. \looseness=-1
\qianwen{We hope to see more novel ML techniques that are tailored for the unique needs in visualizations, such as the differences between chart images and natural images~\cite{Kembhavi2016diagram, chen2019towards}, the differences between user interactions with visualizations and general user interactions~\cite{Dimara2020What}.}

\vspace{0.5em}\noindent
\textbf{Human-Machine Collaboration:}
% Most ML4VIS studies propose end-to-end solutions without the involvement of humans.
Most ML4VIS studies treat ML as a black box without the involvement of humans.
% However, keeping human in the loop is a main characteristic and advantage of visualization.
Take \emph{\revision{Data-VIS Mapping}} as an example.
ML directly generates suitable visualizations from the given data in an end-to-end manner~\cite{data2vis2019CGA,vizML2019CHI}.
However, visualization is by nature a human-centric field: visual design often relies on the creativity of designers; visual analysis usually depends on the knowledge and experience of domain experts. 
Keeping humans in the loop is crucial for the success of visualization.
% From the aspect of visual analysis,
% some tasks are complex, ambiguous, and can hardly be solve through an end-to-end approach. 
% From the aspect of visual design, 
% design is a highly creative and subjective process.
More importantly, a perfect ML model can rarely be obtained limited by the quality of training data and the ambiguity of the problem.
Unsatisfactory model results will decrease user trust in ML4VIS and hinder the wide adoption of ML4VIS.
Take \emph{VIS Interaction} as an example, automated interaction refinement can confuse users when the prediction is at odds with the user's expectation~\cite{FanH2018fast,chen2019lassonet}.

Considering the possibility of unsatisfactory model results, 
many ML4VIS studies allow user participation by supporting post-hoc refinement, such as adjusting the automatically selected nodes~\cite{chen2019lassonet} and modifying the auto-generated visualizations~\cite{cui2019text}.
Recently, initial exploration in human-machine collaboration has been conducted~\cite{lekschas2020peax, tang2020plotthread}.
For example,
PlotThread \cite{tang2020plotthread} enables the ML agent and the human designer to work in the same canvas and modify the layout of a storyline collaboratively. 
However, it requires further exploration of how to support a close human-machine collaboration where human users are able to interpret, modify, and improve the ML.
\qianwen{
We envision that a close human-machine collaboration in ML4VIS can be achieved through explainable and interactive ML, which not only increases user trust by providing transparent predictions~\cite{chatzimparmpas2020state} but also improves ML performances by utilizing user feedback~\cite{Ming2019prototypes}.
}

\vspace{0.5em}\noindent
\textbf{Towards User-friendly ML4VIS:}
The employment of ML not only provides opportunities but also poses new challenges in designing visualizations.
For example, using ML to automatically refine user interaction or generate visualizations can violate the design principle of \textit{``minimize unexpected changes''} \cite{nielsen1990heuristic, Amershi2019guidelines}.
% discussed the challenges that the application of ML will pose towards a user friendly system.

Some ML4VIS studies have discussed the usability issues of ML4VIS systems and proposed design suggestions in discussion and future work. 
For example, Wang et al.~\cite{wang2019deepdrawing} suggested that self-exploring should be supported for expert users in automated graph drawing.
When evaluating their proposed natural language interface for visual analysis,
Yu et al.~\cite{yu2019flowsense} found that users were often confused about why their queries were rejected and did not know how to modify the queries.
Yu et al. suggested that possible corrections of rejected queries should be provided to improve the usability of the system.

While these design suggestions are insightful, they are scattered among different papers.
A systematic summary of these design guidelines is still missing.
More importantly, 
systematic cognitive studies are required to help designers better understand user behaviors and expectations in this new ML4VIS scenario.

% More users evaluation are needed for ML4VIS.

% The design considerations of ML4VIS, design ML4VIS from the users' perspective.
% to helps visualization developers, end users, general audience, who usually lack a high level of expertise in ML.

% the NLP for visual analysis.
% "Toward Interface Defaults for Vague Modifiers in Natural Language Interfaces for Visual Analysis"
% "Improving the Robustness of Scagnostics"

% Not replacing Humans, but promise a good opportunity for human-machine Collaboration

% % 1. the design of ML methods involves human knowledge about the visualizaiton, cognitive perception. e.g., design the loss function (optimization goal) when training a model

% % 2.we need user study results to help us understand what is the best way to design, implement ML4VIS

\section{Discussion}
\qianwen{In this section, we discuss the restrictions of ML4VIS and the limitations of this survey.
The discussion of ML restrictions complements the review of ML4VIS, aiming to provide visualization researchers and practitioners a comprehensive introduction of ML4VIS.
The limitations of this survey highlight several promising directions for future surveys to further deepen our understanding about ML4VIS.}

\subsection{ML is Not a Panacea}
In spite of the great success of ML4VIS, 
we still want to emphasize that ML is not the only technical solution towards intelligent visualization.
ML-based methods are good at recognizing complicated patterns from a huge amount of data.
However, when the pattern is simple or when training data is hard to obtain, ML may not be the optimal solution.
Previous studies~\cite{savvides2019significance, zhao2017safe} have demonstrated the effectiveness of non-ML-based methods, such as expert-defined rules and statistic summaries, in solving visualization problems, 
especially when the training data is hard to collect, when the problem space is sparse, or when the solution can be easily described through rules.
For instance,
ML techniques are widely used to understand visualizations in the form of bitmap images~\cite{chen2019towards, savva2011revision}.
But if the given input is an SVG file, a traverse of the SVG tree can effectively deconstruct the visualization and extract the content \cite{harper2014deconstructing,harper2017converting}, saving the great efforts to collect training data and the high computation cost to train an ML model.
Another example is FlowSense~\cite{yu2019flowsense}, which uses a combination of pre-defined rules and ML methods to capture natural language input patterns due to the lack of labeled examples and computational resources. 
While we hope this survey can demonstrate the effectiveness of ML in visualizations and promote more ML4VIS studies, visualization researchers should also pay attention to the unique advantages of non-ML-based methods and apply them when applicable.

% the unique advantages of non-ML-based methods
% Therefore, we hope visualization researchers should pay attention to non-ML-based methods and apply them when applicable.

% Previous studies have demonstrated that expert defined rules can simply 

\subsection{Limitations \& Future Work}
% What is the difference between Optimization and Machine Learning and why should you care (https://towardsdatascience.com/what-is-the-difference-between-optimization-and-deep-learning-and-why-should-you-care-e4dc7c2494fe)
% "The first difference is the metric function. "
% "The second important difference is the data. In optimization, we care only about the data in hand. We know that finding the maximum value will be the best solution to our problem. In Deep Learning, we mostly care about generalization"
This survey comes with certain limitations due to the adopted approaches. 
First, we only investigate technique \& application papers in this survey. Apart from technique \& application papers, evaluation papers are also important for the field of ML4VIS.
For example, Hearst et al. \cite{hearst2019toward} conducted a crowdsourcing study with 274 participants to understand how to design natural language interfaces for visual analytics systems. 
A review of these evaluation papers can help us understand how users perceive ML4VIS applications and guide the design of ML4VIS.
%
% Second, in this survey, we mainly investigate one facet of ML (\ie, the type of ML tasks) in order to answer \textit{``what ML capabilities can be provided for ML4VIS?''} without diving into the algorithm details.
% Investigating other facets of ML can also help us better understand ML4VIS studies from other aspects.
% Other interesting facets of ML includes the types of input (\eg, graph, image, sequence) and the types of models (\eg, Markov model, linear regression).
% These investigations
% can provide more detailed instructions for choosing proper ML models in ML4VIS.
% For example, when the input is image, convolution neural networks are good options for solving the problem.
% Many ML textbooks are organized by the types of model input~\cite{goodfellow2016deep}.
Second, we summarize the main purposes of employing ML in different visualization processes based on our analysis of the collected papers.
We believe these analysis can provide useful insights. 
However, it is possible that there are other under-explored areas of employing ML.
ML4VIS is still an ongoing research field, we expect more studies to be conducted to expand our understanding about it.
% Nevertheless, we are confident that the summarized seven visualization processes are stable across different scenarios, given their consistency with the existing theoretical models.

\section{Conclusion}
In this paper, we survey \paperNum{} papers to understand the current practices in ML4VIS research, \ie, employing ML techniques for solving problems related to data visualization.
% This survey aims to answer
Guided by two motivating questions: \textit{``what visualization processes are assisted by ML?''} and \textit{``what ML capabilities are used for visualization?''},
we summarize seven main visualization processes that are benefiting from the application of ML techniques.
% For each visualization process, we summarize the main goals of employing ML with representative examples to illustrate the capabilities of ML techniques.
% To map out the role of ML4VIS in general visualizations,
% we referred to existing visualization models and organize the seven visualization processes in an ML4VIS pipeline.
The seven visualization processes are also aligned with the learning tasks in ML to reveal how the needs in visualization can be formed as ML tasks.
An ML4VIS pipeline is also proposed to organize the seven visualization processes and map out the role of ML4VIS in general visualizations.
We further discuss the current practices and future research opportunities in ML4VIS based on our analysis of the collected studies. 
% ML4VIS is still an ongoing research field.
We believe this survey can provide useful insights into the field of ML4VIS and promote future studies.

\ifCLASSOPTIONcompsoc
  \section*{Acknowledgments}
\else
  \section*{Acknowledgment}
\fi
 
%  This research was supported in part by Hong Kong Theme-based Research Scheme grant T41-709/17N. 
This research was supported in part by Hong Kong Theme-based Research Scheme grant T41-709/17N and the Singapore Ministry of Education (MOE) Academic Research Fund (AcRF) Tier 1 grant 20-C220-SMU-011.
We would like to thank all the anonymous reviewers for their constructive comments.


\ifCLASSOPTIONcaptionsoff
  \newpage
\fi



%\bibliographystyle{IEEEtran}

% \bibliographystyle{abbrv}
\bibliographystyle{IEEEtran}

\bibliography{ref}


\begin{IEEEbiography}
[{\includegraphics[width=1.1in, keepaspectratio]{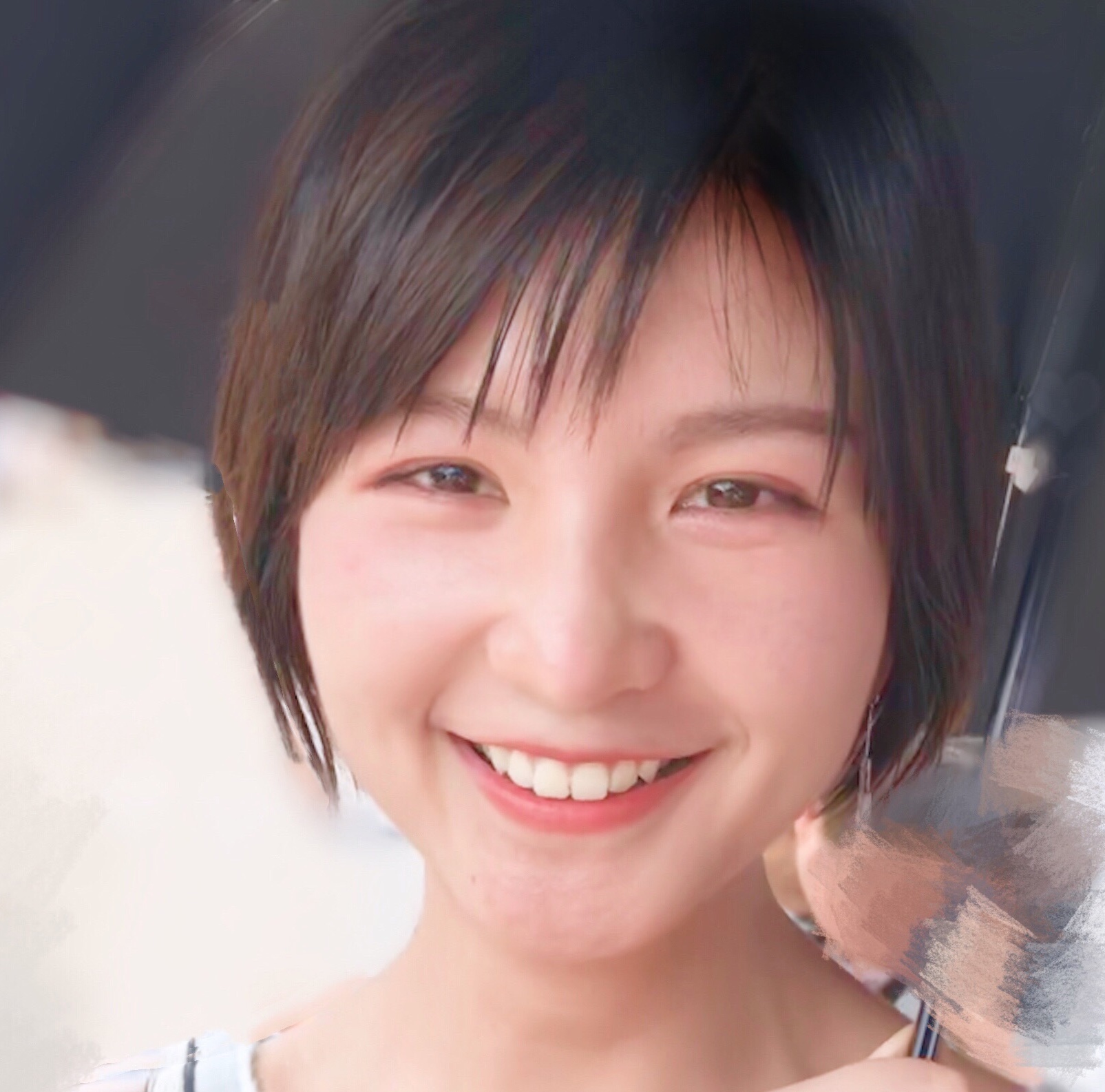}}]
{Qianwen Wang}
is a PostDoc researcher at Harvard University. Her research interests include visual analytics, explainable machine learning, and narrative visualization. Her work strives to facilitate the communication between humans and machine learning models through interactive visualization. 
She obtained her Ph.D in Hong Kong University of Science and Technology and her BS degree from Xi’an Jiaotong University. Please refer to \url{http://wangqianwen0418.github.io} for more details.
\end{IEEEbiography}

\begin{IEEEbiography}[{\includegraphics[width=1.1in,height=1.4in,clip,keepaspectratio]{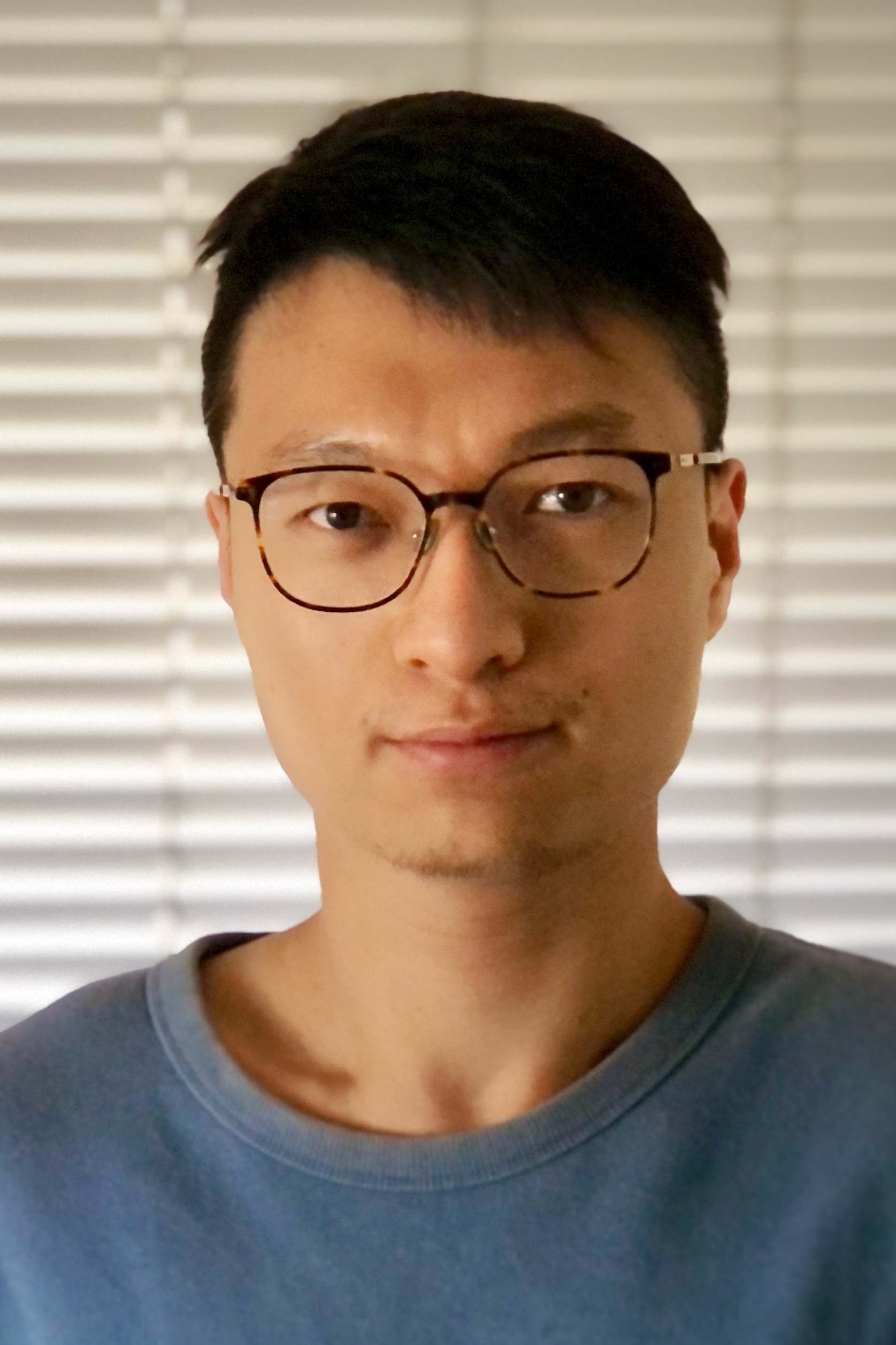}}]{Zhutian Chen}
is a PostDoc researcher at University of California San Diego. His interests are in Information Visualization, Human-Computer Interaction, and Augmented Reality. He obtained his Ph.D. in Computer Science
from Hong Kong University of Science and Technology in 2020. More details can be found at \url{https://chenzhutian.org}.
\end{IEEEbiography}

%\newpage

\begin{IEEEbiography}[{\includegraphics[width=1.1in,height=1.4in,clip,keepaspectratio]{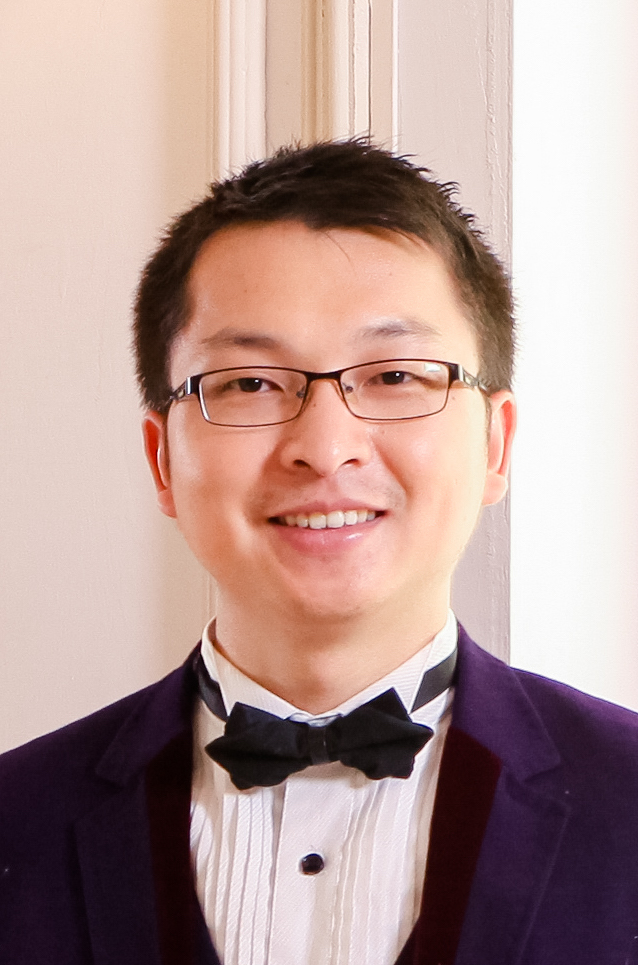}}]{Yong Wang}
is currently an assistant professor in School of Information Systems at Singapore Management University. His research interests include data visualization, visual analytics and explainable machine learning.
He obtained his Ph.D. in Computer Science
% and Engineering 
from Hong Kong University of Science and Technology in 2018. He received his B.E. and M.E. from Harbin Institute of Technology and Huazhong University of Science and Technology, respectively.
For more details, please refer to \url{http://yong-wang.org}.
\end{IEEEbiography}

\begin{IEEEbiography}[{\includegraphics[width=1in,height=1.25in,clip,keepaspectratio]{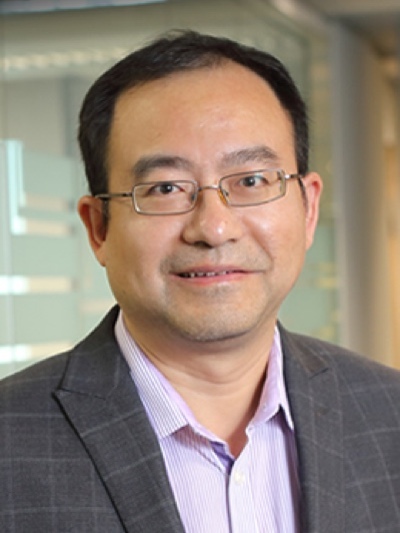}}]{Huamin Qu} is a professor in the Department of Computer Science and Engineering (CSE) at the Hong Kong University of Science and Technology (HKUST) and also the director of the interdisciplinary program office (IPO) of HKUST. He obtained a BS in Mathematics from Xi'an Jiaotong University, China, an MS and a PhD in Computer Science from the Stony Brook University. His main research interests are in visualization and human-computer interaction, with focuses on urban informatics, social network analysis, E-learning, text visualization, and explainable artificial intelligence (XAI).
\end{IEEEbiography}


%\vfill


\end{document}